\pdfoutput=1 %arXiv admin added this line
\documentclass[graybox]{svmult}

\usepackage{mathptmx}       % selects Times Roman as basic font
\usepackage{helvet}         % selects Helvetica as sans-serif font
\usepackage{courier}        % selects Courier as typewriter font
\usepackage{type1cm}        % activate if the above 3 fonts are
\usepackage{makeidx}         % allows index generation
\usepackage{graphicx}        % standard LaTeX graphics tool
\usepackage{multicol}        % used for the two-column index
\usepackage[bottom]{footmisc}% places footnotes at page bottom
\usepackage{bm, amsmath, amssymb}% bold math
\makeindex             % used for the subject index
\begin{document}

\newcommand{\omz}{\omega_z}
\newcommand{\gcqed}{g}
\newcommand{\gpm}{g_\pm}
\newcommand{\gplus}{g_+}
\newcommand{\gminus}{g_-}
\newcommand{\gom}{g_{\mathrm{om}}}
\newcommand{\Perot}{P\'{e}rot}
\newcommand{\dca}{\Delta_{\mathrm{ca}}}
\newcommand{\Dca}{\Delta_{\mathrm{ca}}}

\newcommand{\hami}{H}
\newcommand{\hmech}{H_{\mathrm{mech}}}
\newcommand{\Zcm}{Z_{\mathrm{cm}}}
\newcommand{\Zho}{Z_{\mathrm{ho}}}
\newcommand{\hbath}{\hami_{\mathrm{bath}}}
\newcommand{\Zurich}{Z\"{u}rich}
\newcommand{\hamiom}{\hami_{\mathrm{om}}}
\newcommand{\snn}{S_{nn}}
\newcommand{\nbar}{\bar{n}}
\newcommand{\Datom}{D_{\mathrm{atom}}}
\newcommand{\Dmode}{D_{\mathrm{mode}}}

\newcommand{\bcos}{\hat{b}_{\mathrm{cos}}}
\newcommand{\bsin}{\hat{b}_{\mathrm{sin}}}

\title*{Cavity optomechanics with cold atoms}
\author{Dan M. Stamper-Kurn}
\institute{Dan M. Stamper-Kurn \at Department of Physics, University of California, Berkeley, CA 94720, USA, and \\ Materials Sciences Division, Lawrence Berkeley National Laboratory, Berkeley, CA 94720, USA, \email{dmsk@berkeley.edu}}

\maketitle

\abstract{The mechanical influence on objects due to their interaction with light has been a central topic in atomic physics for decades.  Thus, not surprisingly, one finds that many concepts developed to describe cavity optomechanical systems with solid-state mechanical oscillators have also been developed in a parallel stream of scientific literature pertaining to cold atomic physics.  In this Chapter, I describe several of these ideas from atomic physics, including optical methods for detecting quantum states of single cold atoms and atomic ensembles, motional effects within single-atom cavity quantum electrodynamics, and collective optical effects such as superradiant Rayleigh scattering and cavity cooling of atomic ensembles.  Against this background, I present several experimental realizations of cavity optomechanics in which an atomic ensemble serves as the mechanical element.  These are divided between systems driven either by sending light onto the cavity input mirrors (``cavity pumped''), or by sending light onto the atomic ensemble (``side pumped'').  The cavity-pumped systems clearly exhibit the key phenomena of cavity optomechanical systems, including cavity-aided position sensing, coherent back action effects such as the optical spring and cavity cooling, and optomechanical bistability; several of these effects have been detected not only for linear but also for quadratic optomechanical coupling.  The extreme isolation of the atomic ensemble from mechanical disturbances, and its strong polarizability near the atomic resonance frequency, allow these optomechanical systems to be highly sensitive to quantum radiation pressure fluctuations.  I describe several ways in which these fluctuations are observed experimentally.  I conclude by considering the side-pumped cavity experiments in terms of cavity optomechanics, complementing recent treatments of these systems in terms of condensed-matter physics concepts such as quantum phase transitions and supersolidity.}

\section{Introduction}

During the 1980's and 1990's, a major fraction of atomic physics research was focused on the mechanical effects of light-atom interactions (discussed nicely in the 1997 Nobel lectures \cite{chu98nob,cohe98nob,phil98nob}).  This research area, the roots of which reach back much earlier, might now be called ``atomic optomechanics,''   providing a second intellectual background, parallel to the studies related to gravity wave detection and quantum measurement limits, for present-day investigations of optomechanical interactions with massive solid-state objects.

Beyond this conceptual confluence, research on cavity optomechanics and on light-atom interactions has been united in recent years by experiments in which solid-state objects such as mirrors and membranes inside a cavity are replaced with gas-phase mechanical objects -- non-degenerate ensembles of atoms, Bose-Einstein condensates, and even single trapped ions.  These atomic cavity  optomechanical systems circumvent the difficulties of preparing mechanical systems in the quantum regime by borrowing the methods of laser- and evaporative-cooling developed for the study of quantum gases. This capability has allowed experimentalists to explore quantum properties of both the ``opto'' \cite{broo11pond} and ``mechanical'' \cite{brah12sideband} portions of cavity optomechanical systems, to achieve new regimes of optomechanical coupling, and to explore similarities between optomechanics and paradigmatic many-body Hamiltonians \cite{baum10dicke}.

In this Chapter, I attempt to summarize research on the optomechanics of atoms and atomic ensembles within optical cavities.  The discussion begins with a review of basic single-atom optomechanical effects that are relevant to the ensuing discussion.  I then summarize the single-atom-based view of cavity optomechanics that supplemented the study of single-atom cavity quantum electrodynamics (cQED) by considering mechanical effects such as diffusion, cooling and trapping.  In that work, the motion of a single atom clearly represents a single mode of motion -- or perhaps three modes if not only the axial motion is considered -- and the connection to solid-state cavity optomechanics is readily apparent.  In Sec.\ \ref{sec:continuousmedia}, the discussion turns from single atoms to atomic ensembles.  I describe several investigations of optomechanical effects in continuous atomic media, highlighting situations were a small number of mechanical modes of the ensemble are at play; these include collective Rayleigh and Raman scattering from cold atomic gases, and the realizations of these phenomena within optical cavities.  In Sections \ref{sec:cavityoptomechanics} and \ref{sec:sidepumped}, these precedents are combined to describe the realization of quantum cavity optomechanics with both spatially extended and confined cold-atom ensembles, summarizing recent research results and highlighting the new regimes accessible in these systems.

\section{Basics of light-atom interactions}

The mechanical interactions between single atoms and an optical field may be differentiated according to whether the light-atom interaction is dissipative or dispersive.  Dissipative interactions cause light to be absorbed, meaning that light power is transmitted with sub-unity efficiency due to the fact that photons are scattered out of the incident light beam.  The scattered photons are emitted at random directions, according to an angular probability distribution function determined by the dipole emission pattern,  and, disregarding antibunching over the very short lifetime of the atomic excited state,  at random times.  Dispersive interactions cause light to be phase shifted.  As this phase shift is spatially varying, according to the position of the atom(s), the dispersive interaction also causes optical power to be redistributed, e.g.\ among the wavevectors of the incident light, in a deterministic manner.  The dissipative and dispersive interactions can be treated on equal footing by considering the imaginary and real parts of the complex optical susceptibility (or, equivalently, the index of refraction), respectively.  The two types of interactions can also be described as either spontaneous or stimulated emission.

The discussion above pertains both to Rayleigh (or Bragg) scattering, wherein an atom scatters photons and returns to the same internal state, and to Raman scattering, wherein light scattering shuffles an atom between internal states.  In the former case, it is helpful to reduce the complexity of the atom to just two internal levels, the ground and excited internal states.  This two-level atom approximation is valid when Raman scattering pathways are suppressed either by selection rules and the use of suitably polarized light, or when the light is sufficiently detuned from atomic resonances.

These two aspects of light-atom interactions lead to two types of optical forces on the atom.  The mean force due to dissipative interactions is known as the scattering force or as radiation pressure, whereas that due to dispersive interactions is known as the optical dipole or gradient force.  A unified derivation of both types of forces, making use of the optical Bloch equations to trace the internal-state evolution of an atom in a spatially inhomogeneous optical electric field $\mathbf{E}$, shows the scattering force to be proportional to the spontaneous scattering rate and the optical dipole force to be proportional to the Stark shift $\langle \mathbf{d} \cdot \mathbf{E}\rangle$, where $\mathbf{d}$ is the electric dipole operator \cite{gord80}.  In both cases, the forces arise from the redistribution of the momentum of the electromagnetic field due to the atom.  The optical dipole force can also be envisioned via the dressed-state picture, in which an atom driven by a monochromatic laser field is treated in a time-varying frame that rotates with the laser frequency.  In the dressed-atom approach, the Hamiltonian is time-independent, with eigenstates composed of products of atomic and optical quantum states.  Approximating the atomic motion to be slow compared to the internal-state dynamics, the local eigenenergies of these dressed states define effective optical dipole potentials -- the ac Stark shifts -- from which optical dipole forces are derived \cite{dali85}.  Importantly, both treatments can be applied to describe velocity-dependent forces, which appear for both the scattering and the optical dipole force, and the effects of atomic saturation at high light intensity.

\begin{figure}[t]
  \begin{center}\includegraphics[width=7 cm]{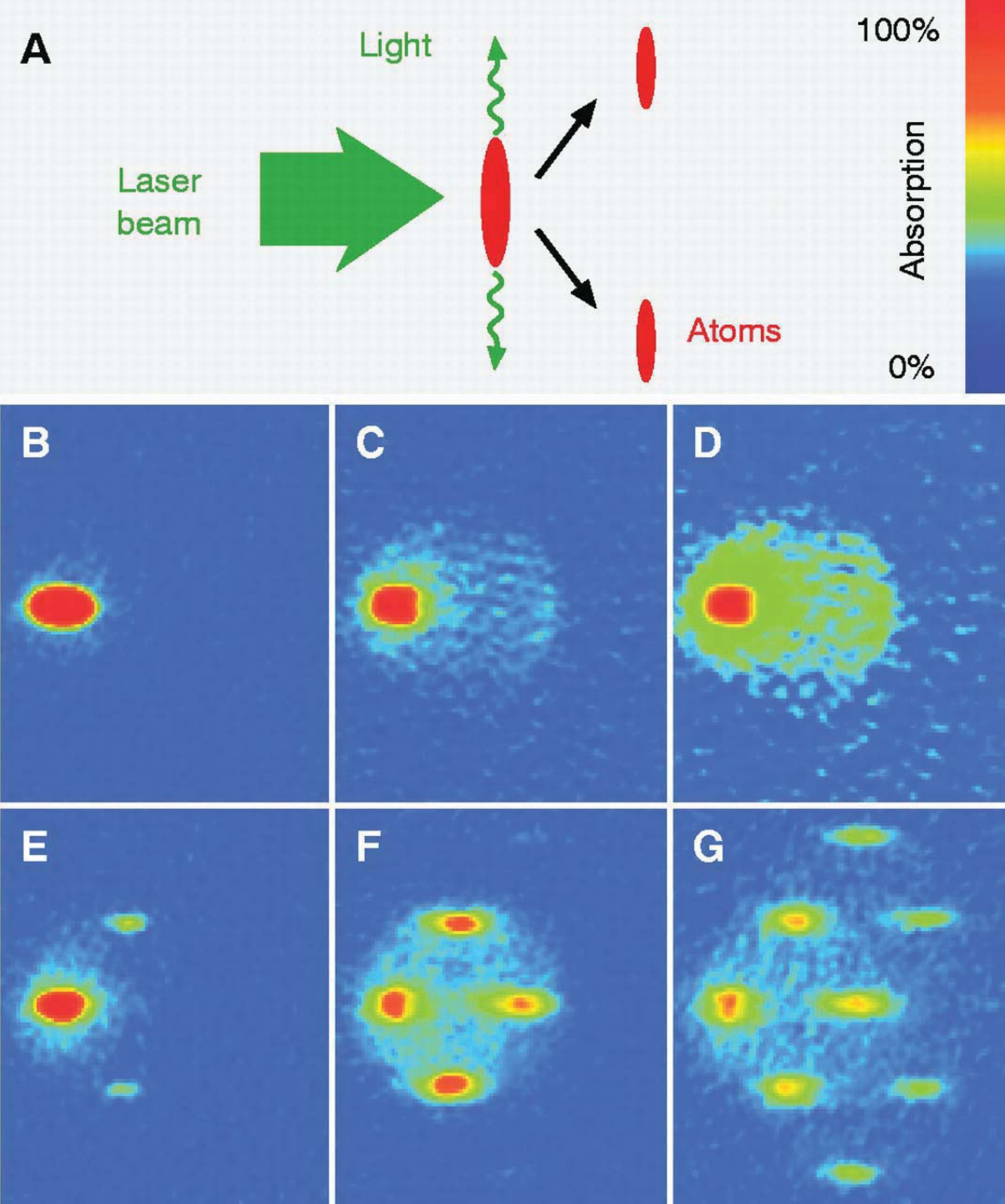}\\
  \caption{Mechanical effects on atoms of single photon scattering.  (A) A Bose-Einstein condensate is illumined with off-resonant light.  (B-D) Choosing a  polarization that inhibits light scattering down the long axis of the condensate suppresses collective scattering.  After time of flight, a halo of scattered atoms is observed, indicating the momentum transferred to atoms by single photon scattering distributed according to a dipole emission pattern.  (E-G) Superradiant Rayleigh scattering: Allowing emission along the condensate axis allows for Brillouin instability.  The coherent momentum populations emitted by light scattering indicate the wavevectors of the instability.  At high probe fluence (increasing from left to right images), high-order Brillouin instability generates coherent populations at multiples of the recoil momentum.  Figure reproduced from Ref.\ \cite{inou99super2}.} \label{fig:SRfigure}
  \end{center}
\end{figure}

Owing to their small mass, atoms are strongly influenced by radiative forces, in contrast with solid-state mechanical systems for which radiative forces may be only a small perturbation atop other influences.  The scattering force can be as large as $\hbar k \gamma$, where $k$ is the photon wavevector and $\gamma$ is the excited-state half-linewidth.  For $^{87}$Rb and light at the 780-nm-wavelength atomic resonance, this maximum force yields an acceleration of $10^5 \, \mbox{m}/\mbox{s}^2$.  The optical dipole force is not limited by saturation, and can thus be even larger \cite{dali85}.  As illustrated in Fig.\ \ref{fig:SRfigure}, optomechanical effects from the scattering of just a single photon are easily discernible given that the initial momentum of a neutral atom can be brought below the single-photon recoil, $\hbar k$, by cooling a gas to well below the recoil temperature $T = \hbar^2 k^2 / 2 m k_B$ (where $m$ is the atomic mass).

The optical forces on an atom fluctuate, leading to the diffusion of its momentum. The diffusion can be associated with two distinct processes. One process is the fluctuation of the atomic dipole, arising due to the quantum mechanical light-atom dynamics, which leads to fluctuations of the optical dipole force in the presence of an electric field gradient.  A second process is the interaction of the atomic dipole with quantum fluctuations of the electromagnetic field.  In the treatment of Gordon and Ashkin \cite{gord80}, for example, the atom interacts with quantum fluctuations of the electromagnetic field in free space, which are local both in time and space, i.e., the shot-noise spectrum of the light field is white and the atom interacts with fluctuations of the optical field at all incident angles.  This second process is modified within optical cavities as the quantum fluctuations of the optical field become modified by the cavity spectrum.  The atom responds to the vacuum noise of the electromagnetic field by spontaneously emitting photons, and recoiling from each emission by a momentum $\hbar k$ directed counter to the photon emission direction.  In the language of optomechanics, such recoil heating may be called the quantum fluctuations of the radiation pressure force, or radiation pressure shot noise.  While such noise is challenging to observe for solid-state objects exposed to light, it is observed routinely in laser cooling experiments.

\section{Optomechanics of single atoms in cavities}
\label{sec:1atomcqed}

The interaction between a single atom and light within an electromagnetic cavity, described by the theory of cQED, has been studied by atomic and optical physicists for decades.  The cavity amplifies the influence of a single optical mode on the internal (i.e.\ electronic) dynamics of the atom so that coherent processes, such as the stimulated re-emission of photons into the cavity mode, can dominate the dissipative processes which typify light-atom interactions in free space.  The deterministic exchange of energy between the electronic excitations of the atom and photons of the cavity field offers the prospect for quantum devices such as quantum memory registers, entangling quantum gates, and, for cascaded atom-cavity systems, quantum networks.

Early experiments on cQED in the optical domain were performed with high-velocity atomic beams transiting the optical resonator, so that the number and position of atoms within the cavity field was uncertain.  While this condition was good enough to illustrate basic cQED phenomena, more ambitious experiments required that the nuisance of atomic motion within the cavity be controlled.

\subsection{Sensing the position of a single atom}

The goal of position sensing for single atoms has been pursued both for atoms in free space and within optical resonators.  Several free-space methods reminiscent of magnetic resonance imaging have been developed in which inhomogeneous magnetic or optical fields lead to strong spatial variation of optical absorption lines \cite{salo87channeling,thyw05encoding} or (narrower) Raman transitions \cite{thom89uncertainty,kunz94}.  In the former case, the atomic position distribution is inferred by the shape of the optical absorption spectrum.
%, for example, to demonstrate the channeling of atoms by a standing wave of light \cite{salo87channeling} or to measure the equipotential contours of an optical trap \cite{bran08}.
In the latter, the atomic position becomes encoded in the atomic internal state which, in turn, can be detected efficiently.  This Raman resonance imaging method was used to measure atomic distributions in beams with sub-optical-wavelength (200 nm) resolution \cite{stok91moving,gard93suboptical}.

These free-space schemes provided some of the motivation for cavity-based schemes of single-atom imaging. Adopting the two-level atom approximation introduced above, the interaction of a single atom with a single cavity mode is quantified by the vacuum Rabi frequency, $2 \gcqed(\mathbf{r}) \propto \mathbf{d} \cdot \mathbf{E}(\mathbf{r})$, which is the rate at which photons are cyclically emitted and reabsorbed by an excited atom placed in an empty cavity field.  The interaction strength varies spatially according to the cavity mode's electric field $\mathbf{E}(\mathbf{r})$. In the common example of the TEM$_{00}$ mode of a Fabry-\Perot\ cavity, at the cavity center, the interaction strength varies as
\begin{equation}
\gcqed = g_0 \, e^{- \rho^2 / w_0^2} \sin k z
\end{equation}
where $z$ denotes the position along the cavity axis and $\rho$ the radial distance from that axis, $k$ is the optical wavevector, and $w_0$ is the beam waist radius.  Owing to this spatial dependence, the position of an atom within the resonator can be inferred by the resonator's optical properties; as in cavity optomechanical sensors employing solid-state objects, the cavity is regarded as an optical sensor of position.

Early works established that measurements on the optical output of a Fabry-\Perot\ cavity lead to a projective measurement of the position of a single atom within the standing-wave intracavity light field.  The authors assumed the atom passes quickly through the light field (e.g.\ originating from a transversely oriented atomic beam), so that mechanical effects of the measurement on the subsequent evolution of the atom, i.e.\ the response to measurement back action, could be neglected.  These works clarified how continuous homodyne measurement of the cavity-emitted light, treated as a quantum-optics measurement process, leads to an ever refined sensing of the atomic position \cite{stor92prl,mart92,stor93pra,herk96localization,mabu96open}.  A feasibility study of this approach, which accounted for realistic experimental parameters, indicated that high-temporal-bandwidth and high-spatial-resolution (below an optical wavelength) was achievable \cite{remp95}.

Subsequent work began taking into account the response of the atom to the continuous cavity-based measurement of its position.  A quantum trajectory simulation showed that, even as the atom was disturbed by the measurement process, the cavity emission, specifically the phase quadrature of the near-cavity-resonant probe field, provided a continual record of its motion \cite{quad95motion}.  Subsequent works clarified that the momentum diffusion experienced by the atom under constant measurement is associated with vacuum fluctuations in the intracavity optical dipole potential experienced by the atom \cite{dunn97}.  This diffusion serves as the back action of a quantum position measurement.

\subsection{Single atom transits and the atom-cavity microscope}

\begin{figure}[t]
\begin{center}
\includegraphics[width= \columnwidth]{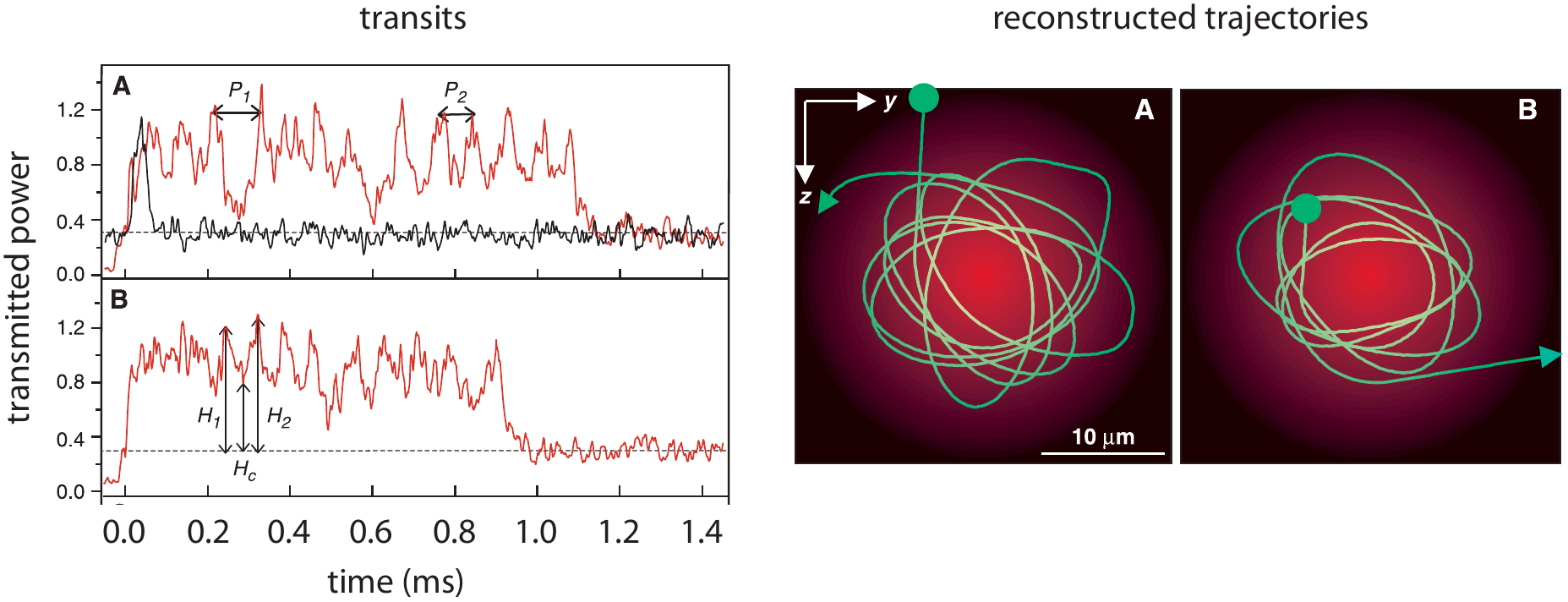}
\end{center}
\caption{Single atom transits and trajectory reconstructions.  (Left) The transmitted power of a cavity probe red-detuned from the empty-cavity resonance is recorded on a heterodyne receiver, revealing two ``transits'' of single atoms passing through the cavity.  The black trace shows reference data for an empty cavity.  (Right) The motion of each atom in the radial plane is reconstructed.  The atom enters the cavity mode, is captured and also perturbed by the optical potential of the cavity probe field, and is finally ejected from the cavity.  Figures are reproduced from Ref.\ \cite{hood00micro}.}
\label{fig:transits}
\end{figure}

Starting in 1996, these theoretical ideas became experimentally relevant with the observation of single-atom ``transits.''  Cold atom sources replaced the atomic beams used previously, so that atoms traversing the cavity would spend enough time -- just a few microseconds at first -- within the cavity mode so that a cavity probe could detect their presence in real time (Fig.\ \ref{fig:transits}) \cite{mabu96}.  Experiments were performed either with probe light resonant with the empty cavity, in which case a passing atom could shift the cavity resonances so that the cavity-transmitted light was variably extinguished, or off-resonant with the empty cavity, in which case the atom would shift the cavity temporarily into resonance and cause the transmitted signal to increase.  The timing of these transits was probabilistic: on each repetition of the experiment, a small number of atoms would be launched toward the cavity, resulting in one or perhaps a few recorded transits.  For medium-input-velocity atoms, the duration of these transits was found to correlate with the incident velocity \cite{muns99singleslow}.  For slower atoms, these transits showed characteristic ripples and wiggles that were taken to represent a real-time record of both internal-state transitions and also center-of-mass dynamics of an atom interacting with the cavity probe field.  These experiments represent first light for atomic cavity optomechanics.

Numerical calculations soon showed that a single atom interacting \emph{mechanically} with a strongly coupled single-mode cavity field is a complex dynamical system.  Here, the single-atom strong coupling parameter is $C = g_0^2 / 2 \gamma \kappa$, also known as the single-atom cooperativity, with $\kappa$ being the half linewidth of the cavity resonance.  The cooperativity quantifies the ratio of the emission rate of an atom into the cavity vs.\ that into free space.  The strong-coupling condition $C \gg 1$ implies that not only the internal dynamics but also the optomechanical effects of light-atom interactions should be influenced strongly by the cavity environment.  Simulations of atomic motion were performed with the atomic position and momentum treated classically, an approximation justified by the very short deBroglie wavelength of the atoms in these first experiments.  For the case of a cavity tuned close to the atomic resonance, which describes most of the single-atom optomechanics work, the simulated atom performed complex dynamics -- hopping randomly between antinodes of the cavity field -- resulting from the strong spatial variation of optical forces and force fluctuations \cite{dohe97motion}.  When the cavity resonance is tuned away from the atomic resonance, atoms within the resonator begin acting dominantly dispersively so that the dynamics more closely resemble those of (dispersive or refractive) solid-state cavity optomechanics.  In this case, the atom was found to experience ``the strongest diffusion [at locations] where the output field contains the most information about the atomic position dynamics.'' \cite{dohe98motional}  In the language of cavity opto-mechanics, the measurement back action upon the atomic momentum is greatest where the measurement sensitivity to the atomic position is strongest.

Backed by these dynamical simulations, the temporal structure seen in the single-atom transit allowed researchers to surmise the real-space trajectory of the atom within the cavity mode (Fig.\ \ref{fig:transits}) \cite{hood00micro,dohe01trapping}.  Interestingly, in contrast to typical cavity optomechanics experiments in which one detects the motion of a cantilever or membrane along the cavity axis, here, the single-atom transit was used to reconstruct the slower motion transverse to the cavity axis, a direction along which the cavity provides only quadratic ($\propto \rho^2$) sensitivity and no sensitivity to angle.

\subsection{Cavity cooling of single atoms and ions}

An important conceptual understanding of how mechanical light-atom interactions are affected by cQED was developed by Ritsch and colleagues, marked by the identification of cavity Doppler cooling and cavity-enhanced diffusive heating \cite{hora97,hech98cool}.  Their work differs from previous derivations in the context of solid-state cavity optomechanics \cite{brag77weakbook} in several respects.  First, their treatment begins with the electric dipole coupling of the atom to the cavity field, explicitly including the atomic excited internal state so as to account for effects of saturation and spontaneous emission.  As such, their work is applicable both to the case of single-atom experiments where the cavity probe field is nearly resonant with the atomic transition and is strongly coupled to the atom, and also to the case of many-atom experiments where the field is far off-resonant.  Their treatment also closely parallels the Gordon and Ashkin treatment of free-space optical forces, cooling and diffusion \cite{gord80}, allowing one to identify the new cavity optomechanical effects as resulting from the electrodynamics of the cavity. While the derivation specifically considers the motion of a two-level atom, the authors also recognize the generality of cavity Doppler cooling, stating that ``As the only requirement on the particle is a strong coupling to the cavity mode, the results should also apply to small molecules or other more complex objects (as, e.g.\, a Bose condensate) with a sufficiently large dipole moment.'' \cite{hora97}

Another valuable element of their work is the analogy between cavity cooling and Sisyphus laser cooling of atoms outside cavities \cite{chu98nob,cohe98nob}.  In free-space Sisyphus cooling, an atom moves in an internal-state- and spatially dependent ac Stark shift produced at the intersection of several near-resonant light fields.  Under proper conditions for cooling, the atom is preferentially optically pumped into the internal state for which the ac Stark potential has a local minimum.  Because optical pumping is not instantaneous, but rather occurs only after a delay $\tau$ related inversely to the spontaneous scattering rate, the atom tends always to ``climb up'' the ac Stark potential energy landscape, always doing work against the optical field.

\begin{figure}[t]
  \begin{center}
  \includegraphics[width= \columnwidth ]{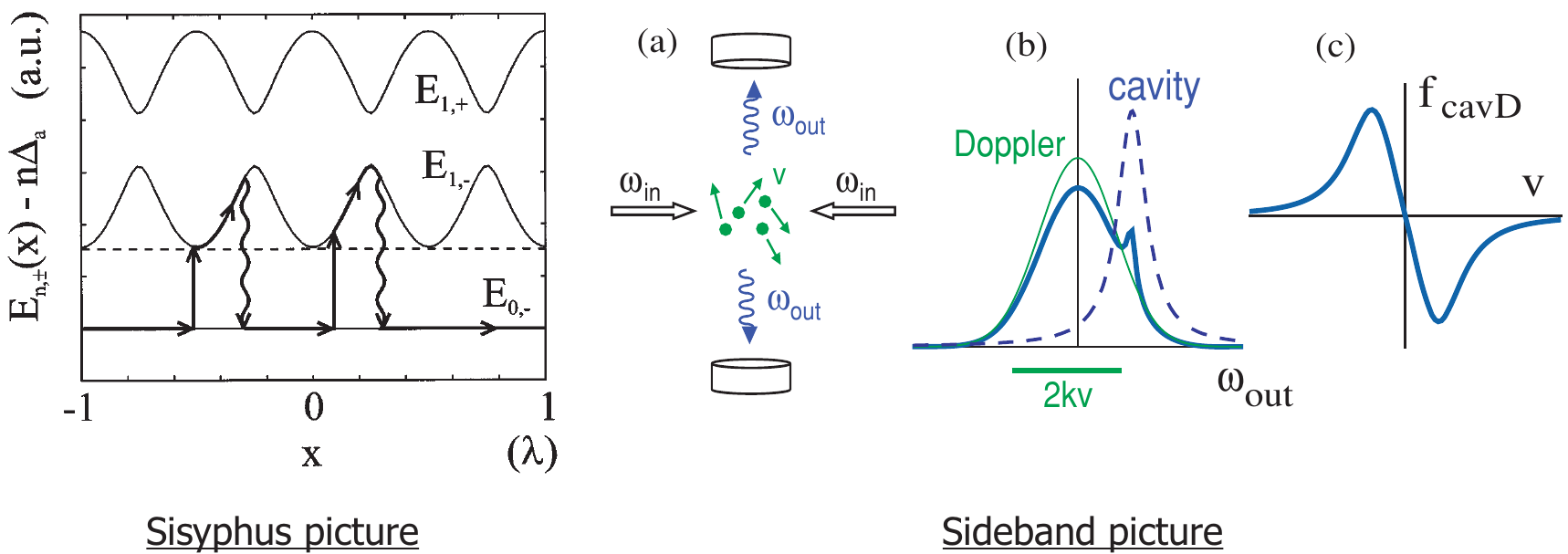}\\
  \end{center}
  \caption{Two conceptual pictures of cavity cooling provided by atomic physics.  Left: In the Sisyphus picture, an atom is excited by a red-detuned cavity probe to the first-excited atom-cavity state, at a position within the cavity where that excited state energy is a local minimum in space.  Moving along the cavity axis, the atom remains in that excited state for a time $\sim \kappa^{-1}$, during which it does work against the optical dipole force, before decaying back down to the spatially uniform ground state.  Right: In a frequency-sideband picture, after an atom within a cavity absorbs a photon from a driving field that is red-detuned from the cavity resonance, it is stimulated by the cavity electromagnetic mode structure to re-emit nearer the cavity resonance frequency, cooling the atom.  Figure (b) shows the overall spectrum of light scattered by an ensemble of rms velocity $v$: a sum of a Doppler profile of transversely scattered light, and a narrow spectrum of cavity-enhanced scattering into the cavity mode(s).  (c) The net energy exchange provides a damping force $f_{\mathrm{cavD}}(v)$ on the atoms.  Figures reproduced from Refs. \cite{hora97} (Left) and \cite{blac05jphysb} (Right).}\label{fig:coolingpicture}
\end{figure}

Similarly, in describing cavity cooling, Ritsch and colleagues consider the lowest lying atom-cavity excited dressed states, with spatially varying energies
\begin{equation}
E_{1,\pm}(\mathbf{r}) = \left( \frac{\omega_a + \omega_c}{2}\right) \pm \sqrt{ \frac{\dca^2}{4} + \gcqed^2(\mathbf{r})}
\label{eq:e1pm}
\end{equation}
Here $\omega_a$ ($\omega_c$) is the atomic (cavity) resonance frequency, and $\dca = \omega_c - \omega_a$.  If the cavity is driven with light tuned at or below the minimum value of $E_{1,-}$, an atom moving, for example, along the cavity axis is preferentially excited at a minimum of the optical-potential energy surface (left side of Fig.\ \ref{fig:coolingpicture}).  The atom then climbs up the potential energy surface before relaxing back to the atom-cavity ground state, either by spontaneous emission or cavity decay, emitting a photon at higher energy than the probe and thus giving up some of its kinetic energy.  Here, the delay time $\tau$ relates inversely to the cavity decay rate.  The larger is the delay time, the longer the atom can work against the optical field and the greater is the cooling force; hence, narrower optical cavities cool the atom more strongly, allowing it to reach lower temperatures.

Vuleti\'{c} and Chu provided an alternative, frequency-space description of cavity cooling \cite{vule00}.  Light scattering involves photon absorption followed by photon emission.  Conventional Doppler cooling makes use of the frequency selectivity of \emph{photon absorption}: Light tuned to the red of an atomic transition is absorbed preferentially by an atom moving counter to the photon wavevector, because the Doppler shift brings such photons closer to the atomic resonance.  The cooling force, depending on the preference for absorbing light that is Doppler shifted toward rather than away from resonance, scales inversely with the linewidth of the transition.  Given also that many scattered photons are needed to slow down an object appreciably, Doppler cooling is effective primarily for simple atoms.

In contrast, cavity Doppler cooling utilizes the frequency selectivity of \emph{photon emission}: An atom inside an optical cavity, illuminated with light to the red of the cavity resonance, will preferentially emit blue-shifted photons, because the electromagnetic density of states is higher nearer to the cavity resonance (right side of Fig.\ \ref{fig:coolingpicture}).  The vacuum-induced stimulation of high-energy photon emission was considered earlier by Mossberg, Lewenstein and Gauthier, although not within the specific context of optical cavities \cite{moss91}.  Vuleti\'{c} and Chu noted that since cavity-induced laser cooling does not rely on narrow atomic resonances, it can be applied to complex objects: molecules which, driven far from their resonances, need not undergo Raman transitions among ro-vibrational states; color centers, via which one can laser-cool solids; and even cantilevers and membranes.

The cavity cooling force is enhanced by making the cavity-stimulated emission stronger with respect to photon emission into free space.  As such, the effectiveness of cavity Doppler cooling can be quantified by the ratio of the photon emission rate into the cavity vs.\ that into free space.  For a single-mode cavity, this ratio is simply the cooperativity $C$; in a multi-mode cavity, this ratio can be increased further \cite{vule01}.

In the demonstrations of cavity cooling of solid-state objects \cite{arci06,giga06cooling,schl06cooling,naik06back}, the moving element is typically held by material supports, compared to which the optical forces affect the element weakly.  Its motion within a narrow temporal frequency band is constrained along a single coordinate; correspondingly, the cavity emission spectrum within a narrow band serves as a (somewhat) direct record of its motion.  The magnitude and width of this spectrum then reflect the damping of the object's motion under steady state conditions.

The motion of single atoms within cavities can be much more complex.  Optical forces from the cavity-cooling light are often the dominant source of confinement, cooling, and diffusive heating. The atom transits the cavity, or resides briefly within it, and may not achieve steady state conditions.  To avoid atomic saturation in strongly coupled cavities, the photon flux through the cavity is kept very low, corresponding to an average cavity photon number on the order of unity.  The data collected from such experiments is thus noisy and transitory.

For these reasons, the early demonstrations of cavity cooling of single atoms relied on subtle data analysis and comparisons to complex numerical models.  For instance, M\"{u}nstermann \emph{et al.}\ examined the cavity spectrum produced with atoms transiting the cavity, and interpreted the spectrum as evidence for cavity-accentuated momentum diffusion and cavity cooling \cite{muns99dyn}.  In later experiments, cavity probes were applied to atoms trapped briefly within the cavity, and a slight increase in trapping lifetimes was taken as indicative of cavity cooling \cite{maun04cooling}.

The evidence for cavity Doppler cooling was strengthened once schemes were developed to trap single atoms for as long as several seconds within intracavity optical traps \cite{nuss053d}.  Here, the atom is confined in far-detuned optical traps, providing an essentially conservative trapping potential, while different light fields closer to the atomic resonance produce the dissipative optical forces.  This work also demonstrated a method to attain strong three-dimensional cooling.  By illuminating atoms with light directed transverse to the cavity axis, the strong cavity Doppler cooling force acts in the two-dimensional plane defined by the pump-light and cavity-light propagation axes \cite{vule00}.  In the experiment of Nu{\ss}man \emph{et al.},\ strong cooling in the third spatial dimension is obtained from the strong variation of the atomic resonance frequency in a far-detuned standing-wave optical potential \cite{nuss053d}.

Recently, sideband-resolved (mechanical frequency larger than the cavity linewidth) cavity cooling of a single trapped ion has also been demonstrated \cite{leib09}.  Such cavity-based cooling was proposed already in 1993 by Cirac \emph{et al.}\  \cite{cira93cavitycooling,cira95cavitycooling}, although their interpretation of the cooling mechanism led them to conclude erroneously that such cooling occurs exclusively in the Lamb-Dicke limit, in which the rms size of the atomic center-of-mass distribution is much smaller than the optical wavelength.  Cavity cooling of tightly bound objects was also considered in Ref.\ \cite{vule01}, in which the conditions for ground-state cooling were spelled out; such analysis was later repeated in the context of cavity optomechanics with solid-state oscillators \cite{wils07groundstate,marq07sideband}. In the experiment, the ion was only weakly coupled to the optical cavity ($C \ll 1$).  Thus, even in the resolved sideband regime, the excess diffusive heating due to spontaneous emission outside the cavity prevented the ion from reaching the mechanical ground state.

\subsection{Cavity-enhanced diffusion as measurement back action}

The enhanced position sensitivity provided by a high-finesse cavity must also yield enhanced momentum diffusion, as a form of measurement back action \cite{dunn97,dohe98motional,murc08backaction}.  A quantum optical treatment of force fluctuations, for the case of a two-level atom, shows that the momentum diffusion constant $D$ in the atom-cavity system can be cleanly separated into three components \cite{murr06}:
\begin{equation}
D = (\hbar k)^2 \gamma P_e + | \hbar \nabla \langle \sigma^+ \rangle|^2 \gamma + |\hbar \nabla \langle a \rangle |^2 \kappa
\label{eq:diffusion}
\end{equation}
The first term quantifies diffusion due to atomic spontaneous emission, similar to that in free space, where $P_e$ is the excited state population.  The second two terms quantify diffusion due to fluctuations of the optical dipole force.  The first of these describes the interaction of a fluctuating atomic dipole (with the operator $\sigma^+$ proportional to the dipole moment operator) with the mean field within the cavity; the second describes the interaction of a fluctuating cavity field with the mean atomic dipole.  The strength and spatial dependence of the atom-cavity coupling is implicit in the spatially varying expectation values, $\langle \sigma^+ \rangle$ and $\langle a \rangle$.  In free space, i.e.\ without the oscillating atomic dipole acting back on the driving optical field, the last term in Eq.\ \ref{eq:diffusion} is absent, and one recovers the earlier results of Gordon and Ashkin \cite{gord80}.  In contrast, the diffusion rate inside a cavity is indeed enhanced, reaching a value that is up to $\sim C$ times larger than the free-space rate.

This enhanced diffusion is observed only indirectly in single-atom experiments, gleaned from measurements on normal-mode splitting for atoms transiting \cite{muns99dyn} or trapped within the cavity \cite{maun05} and comparison with numerical Monte Carlo simulations \cite{pupp07fluc}.  More direct quantifications have been obtained from the many-atom cavity optomechanics system, as described in Sec.\ \ref{sec:cavityoptomechanics}.

\subsection{Feedback cooling of a single atom}

Finally, we discuss feedback control and cooling via cavity optomechanics.  In solid-state optomechanics, such active feedback was used to cool mechanical objects by radiation pressure \cite{coha99} before ``passive'' cavity cooling via the natural dynamics of the cavity field was demonstrated.  Feedback control of the motion of atomic ensembles trapped in optical lattice potentials, but outside of a cavity, was implemented by Morrow \emph{et al.} \cite{morr02feedback}.  There, the intensity of the optical lattice beams provided a measurement of the force exerted by the lattice light onto the ensemble, since such forces can be understood as deriving from the coherent exchange of photons between the intersecting lattice beams \cite{rait98revival}.  This signal was fed back to the phase of one of the lattice beams, shifting the lattice potential spatially so as to amplify or damp the atomic motion.

The enhanced, and, ultimately, quantum limited position sensitivity attainable using a high-finesse optical cavity makes such feedback control schemes more powerful, in the sense that they can be effected with higher gain and may even cool a mechanical object into the lowest few quantum states regime of motion.  The cavity enhances light scattering into a single electromagnetic mode, from which information can be extracted efficiently, thus minimizing the information lost to spontaneous emission into very many, usually unmeasured, emission modes.  Several works have explored feedback control of the motion of a single atom within a cavity, in pursuit of the fundamental scientific goal of understanding and optimizing feedback control of an open dynamical quantum system.

Steck \emph{et al.}\ considered controlling the one-dimensional motion along the axis of a Fabry-\Perot\ cavity \cite{stec04qfb,stec06qfb}.  The atoms are assumed trapped within a red-detuned standing-wave potential formed by light that also serves as the measurement probe.  In this case, the sensitivity to the atomic motion is quadratic in the atomic displacement, similar to that realized experimentally by positioning thin membranes \cite{thom08membrane,sank10tunable} or atomic gases \cite{purd10tunable} at the antinodes of the cavity field. The authors consider a ``bang-bang'' feedback scheme, a digital scheme in which the potential strength is toggled between high- and low-curvature settings, with the aim of effecting a Sisyphus-like cooling in which the atom moves away from the trap center against a strong confining force, and then returns to the center against a weaker force.  When the ``bang'' amplitude is sufficiently high, such a scheme is predicted to bring the particle to the lowest parity-even or parity-odd states of the well (the feedback scheme conserves mirror symmetry about the trap center).

A different feedback scheme was considered by Vuleti\'{c} \emph{et al.}\ for slowing down higher-velocity particles traveling along the standing-wave intensity pattern within a driven Fabry-\Perot\ resonator \cite{vule07external}.  This motion generates a cavity-field modulation at a frequency proportional to the particle velocity.  Appropriate frequency-space conditioning of the feedback signal ensures that the cavity field is modulated so as to continually slow down the particle.  Cooling rates and final-temperature limits were assessed taking into account the momentum diffusion of the particle and the limited signal-to-noise ratio of the measurement signal.

\begin{figure}[t]
\begin{center}
\includegraphics[width=11.5cm]{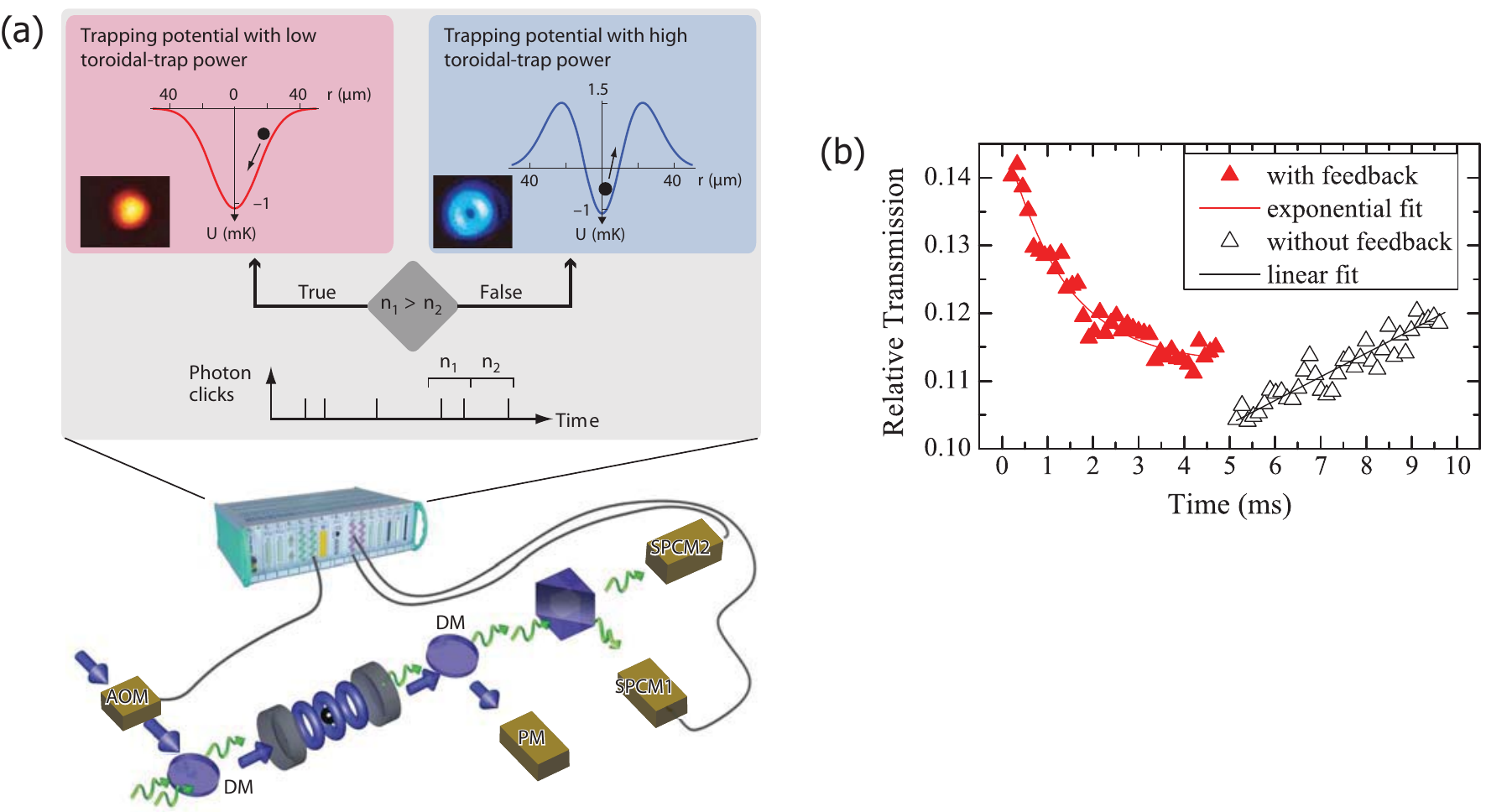}
\end{center}
\caption{Feedback cooling of a single atom within an optical cavity.  (a) The rate of increase of $\rho^2$ is estimated by counting the number of cavity probe photons (using photon counters SPCM1 and SPCM2) in two consecutive time bins.  Based on this difference, the power in a blue-detuned light field is varied between two settings, producing an optical potential on the atom with either low (top left) or high (top right) trap curvature.  (b) With the feedback engaged, the average $\langle \rho^2 \rangle$, proportional to the transmitted probe intensity, is decreased, demonstrating that energy is extracted from the atom by feedback.  With the feedback disengaged, the atom heats up.  Figure (a) reproduced from Ref.\ \cite{kuba09feedback}, and (b) from Ref.\ \cite{koch10feedback}.}
\label{fig:feedback}
\end{figure}

The ``bang-bang'' feedback scheme was successfully implemented by the Rempe group (Fig.\ \ref{fig:feedback}) \cite{kuba09feedback}.  In this implementation, probe photons transmitted by the cavity were counted within consecutive short time bins, so as to estimate whether the atom was moving toward or away from the center of the cavity.  The detection bandwidth was such that only the slower radial oscillatory motion was measured.  The photon-number comparison was used to vary the intensity of light fields that resonated with higher-order transverse modes of the cavity, varying the radial optical confinement.  The feedback provided active cooling, extracting energy from the atom and maintaining trap lifetimes on the order of one second \cite{koch10feedback}.

\section{Optomechanics of continuous atomic media}
\label{sec:continuousmedia}

We have seen how many of the phenomena of cavity optomechanics -- cavity cooling, position sensing, measurement back action, and feedback control -- apply to microscopic, single-atom mechanical oscillators just as they do to macroscopic solid-state objects.  Whether the framework of cavity optomechanics could apply to atomic ensembles, bridging the gap between the microscopic and macroscopic realms, was initially unclear.  For example, treatments of the motion of just two atoms interacting with a single cavity field suggested that their dynamics would be exceedingly complex, subject to cavity-mediated long-range interactions, ``cross-friction'' whereby the motion of one atom is damped due to the motion of the other, and diffusive heating that varies with the positions of each of the particles \cite{fisc01,asbo04twoatoms}.  An animation in Ref.\ \cite{fisc01} displays the simulated dynamics of two atoms dancing wildly within a cavity.

However, several works found evidence that cavity cooling could indeed apply not only to single atoms, but also to atomic ensembles.  For example, the analogy of stochastic cooling of charged-particle beams suggested that dynamic back action of the cavity field could cool thermal fluctuations of an ensemble \cite{vule00}.  Of particular interest was the possibility that the cooling force on an atomic ensemble could be \emph{collectively enhanced}, so that cavity-aided laser cooling could be not only more general, but also more powerful than free-space Doppler cooling \cite{gang99collective,hora01}.

\subsection{Collective cavity cooling and self-organization via Brillouin instability}
\label{sec:brillouin}

\begin{figure}[t]
\begin{center}
\includegraphics[width= \columnwidth]{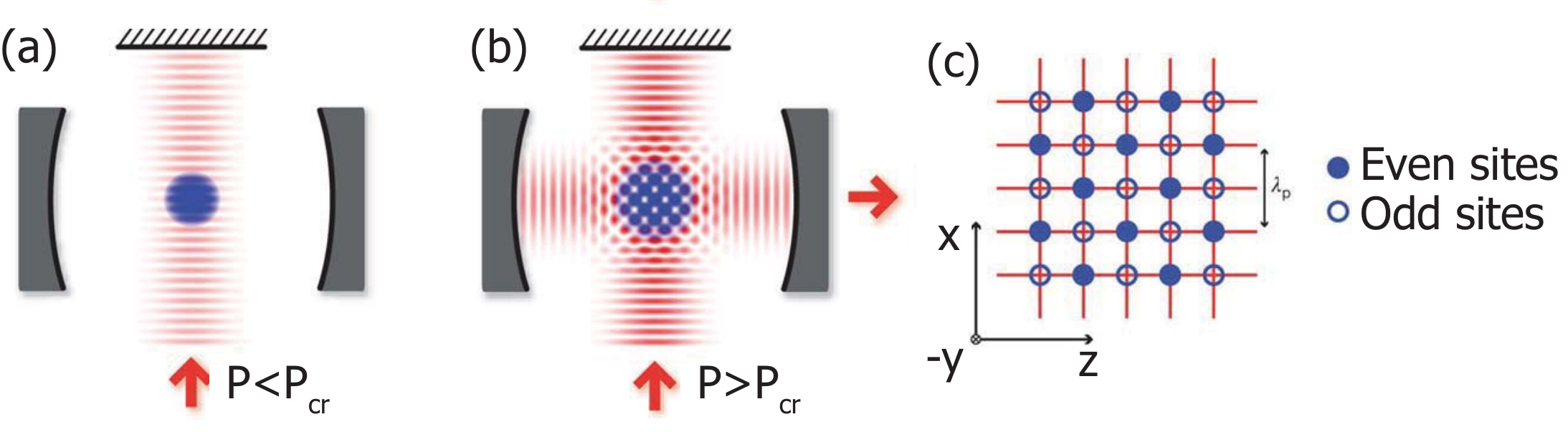}
\end{center}
\caption{(a) A near uniform gas placed within a Fabry-\Perot\ resonator is pumped with a standing-wave of light aligned orthogonal to the cavity axis, and with a frequency near the cavity resonance. (b) Above a threshold pump power, the gas organizes itself spatially so as to scatter pump light efficiently into the resonator.  (c) Maximum collective scattering is achieved in two different spatial patterns of the gas, selecting either the ``even'' or ``odd'' sites of a checkerboard pattern.  Figure adapted from Ref.\ \cite{baum10dicke}.}
\label{fig:evenodd}
\end{figure}

This question was resolved experimentally by the MIT group \cite{blac03}. An atomic gas, pre-cooled to moderate temperatures by conventional laser cooling, was launched at variable velocity within a (multimode) Fabry-\Perot\ cavity, and was then exposed to light aligned transverse to the cavity axis and detuned from the cavity resonance (Fig.\ \ref{fig:evenodd}).  A time-of-flight analysis showed that a large portion of the gas was quickly decelerated, much faster than one would expect just from the theory of single-atom cavity cooling.  The rapid deceleration could be ascribed to collective light scattering by the ensemble into the cavity mode, as confirmed by the fact that although the single-atom cooperativity was low for this setup, the measured ratio of light scattering into the cavity vs. outside the cavity was well above unity.

Such dynamics are now understood to result from the self-organization of the laser-driven atoms into a spatially periodic configuration that enhances the collective scattering by the gas into the cavity field.  That is, the continuous, deformable optomechanical medium organizes itself, in response to cavity-induced forces and damping, into compliance with the cavity field.  This effect was predicted by Domokos and Ritsch, whose theoretical treatment demonstrated the collective nature of the cavity-induced cooling into a checkerboard spatial pattern \cite{domo02collective}.

The process by which a disordered medium spontaneously arranges itself into a collective coherent scattering state is known either as \emph{superradiance} or \emph{lasing}.  The processes are loosely distinguished by whether the buildup of the optical field plays a major (lasing) or minor role (superradiance).

The gain mechanism for such self-organization is a form of Brillouin instability, exemplified first in observations of superradiant light scattering from Bose-Einstein condensates \cite{inou99super2}, and later also from nondegenerate Bose \cite{yosh05super} and Fermi \cite{wang11fermisuperradiance} gases.  In these experiments, the cold gas is trapped in an elongated cigar-shaped geometry.  Exposed to a plane wave of off-resonant pump light, the gas scatters photons, with each scattering event leaving behind a collective momentum excitation which conserves momentum in the scattering process.  This excited residue establishes density modulations in the gas that persist for a time -- very long in Bose-Einstein condensates and much shorter in nondegenerate or Fermi gases -- which depends on the coherence properties of the gas.  While they persist, these modulations preferentially scatter additional pump photons into the same output directions.  This process represents a gain mechanism for density modulations.  The rate constant for this density-grating amplification is proportional to the single-atom Rayleigh scattering rate and the optical depth of the gas along the optical emission direction; hence, for a prolate gas, the highest gain is seen for photons scattered along the long axis of the gas (Fig.\ \ref{fig:SRfigure}). For very strong pump intensities, higher-order superradiant Rayleigh scattering is seen, where the coherently scattered atoms themselves undergo superradiant scattering and produce coherent atom populations at ever higher momenta \cite{inou99super2}.  Regarded in position space, this higher-order process represents the bunching of atoms into a sharp density grating.

Within a cavity, light emission is stimulated not only by the buildup of a material density grating, but also by the occupation of the optical cavity mode.  The aforementioned gain mechanism is then the basis of the ``coherent atom recoil laser'' \cite{boni94CARLnim,boni94CARLpra}.

\begin{figure}[t]
\begin{center}
\includegraphics[width=9cm]{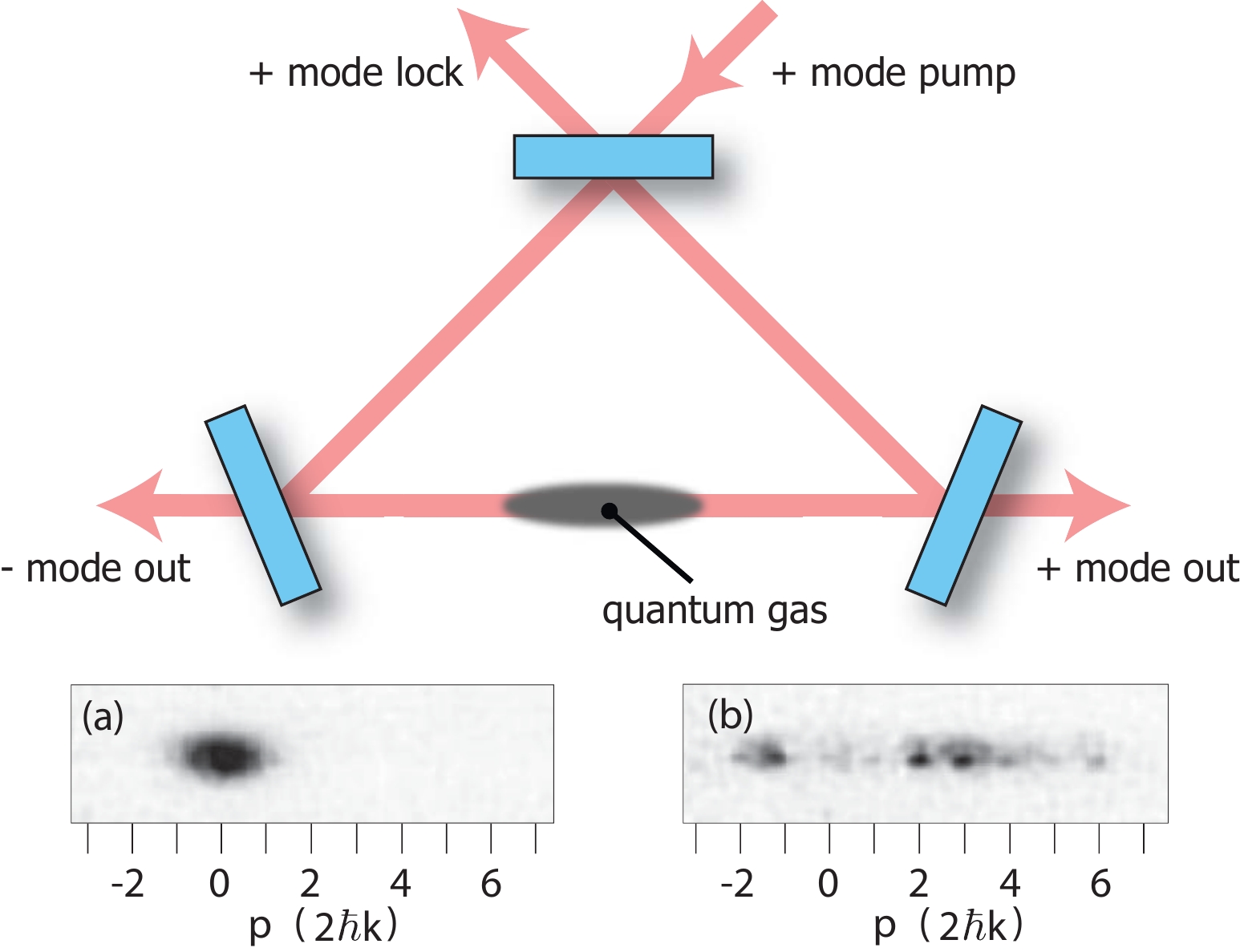}
\end{center}
\caption{Setup for pumped ring-cavity experiments.  A quantum gas is placed within the mode volume of a ring cavity.  The right-going mode (seen by the atoms, labeled with $+$) is pumped by a coherent-state input that is locked onto the cavity resonance.  Monitoring the output of the left-going mode (labeled with $-$) reveals the coherent back scattering of light due to self-organization of the gas.  Time-of-flight distributions of the gas (a) before and (b) after the coherent atom recoil lasing action show that coherent back scattering is accompanied by the distribution of the gas among several coherent momentum populations.  These indicate the arrangement of the atoms into a periodic spatial structure.  Panels (a) and (b) are adapted from Ref.\ \cite{slam07bec}.}
\label{fig:ringcavity}
\end{figure}

The relation between Rayleigh superradiance and the coherent atom recoil laser was studied experimentally by placing a cold atomic gas within a ring cavity (Fig.\ \ref{fig:ringcavity}) \cite{slam07super}.  The atoms-cavity system was pumped through one input port, exciting one running-wave mode of the cavity.  Forward scattering by the atoms induces an atom-number-dependent frequency shift of both running-wave resonances of the cavity.  Back scattering by the atoms couples the two modes. By monitoring the output of the second running-wave mode of the ring resonator, the spontaneous buildup of one-dimensional density modulations in the cold gas was observed.  In addition, a time-of-flight imaging technique indicated the buildup of atom population at multiples of the back-scattering momentum recoil $2 \hbar k$.  By varying the temperature of the gas and parameters of the optical cavity, the relation between the lifetime of the matter-wave density modulations and of the cavity field was tuned between the ``bad-cavity'', or superradiant limit, and the ``good-cavity'' or ``ringing-superradiance'' limits of the coherent atom recoil laser \cite{blac05jphysb,slam07bec}.  Depending on the experimental configuration, the cavity-field and matter-wave amplifications can either favor the same or different forms of the Brillouin instability; for example, a narrow frequency cavity can select the coherent emission of momentum populations different than those expected from free-space superradiant scattering \cite{bux11cavitycontrolled}.

A similar mechanism operates in the MIT (and \Zurich) experiments \cite{blac03,baum10dicke}, where the atoms are pumped by an externally imposed standing-wave optical field with its axis orthogonal to the axis of a Fabry-\Perot\ cavity.  The Brillouin instability organizes the gas into one of two distinct spatial patterns that maximize the collective scattering into the cavity mode, but that are distinguished by the relative phase, either $0$ or $\pi$, between the probe and cavity fields (Fig.\ \ref{fig:evenodd}); the emergence of these patterns was confirmed both by time-of-flight analysis \cite{baum10dicke} and by the observed bistability of the relative optical phase \cite{blac03,baum11breaking}.

In Sec.\ \ref{sec:sidepumped}, we reconsider these experiments in terms of cavity optomechanics.

\section{Quantum cavity optomechanics with cold atoms in a driven Fabry-\Perot\ resonator}
\label{sec:cavityoptomechanics}

As highlighted above, the full dynamics of a few or many particles, the motion of which is coupled by and to the dynamics of an optical resonator, is rich and complex.  It was therefore surprising that a very simple, clarifying, and quantitative treatment of such dynamics, using the paradigm of cavity optomechanics, could emerge.  The key to this simplification was to create experimental situations where the mechanical dynamics of the atomic ensemble is restricted to a small perturbation atop a well characterized initial state.  Such experimental starting points include near-uniform ultracold gases held within the resonator \cite{bren08opto,baum10dicke} and atoms cooled near the ground state of axial motion within many \cite{gupt07nonlinear} or just a few potential wells \cite{purd10tunable} of an intracavity optical lattice.

\subsection{Collective atomic optomechanical response}
\label{sec:collectivevariable}

The derivation of the cavity optomechanics Hamiltonian as an approximation of the dispersive interaction between an atomic ensemble and a single-mode cavity has been outlined in several works \cite{murc08backaction,bren08opto,bott09,schl11cooling}.  For the purpose of this discussion, we focus on forward scattering by the atom, a dispersive two-photon effect wherein an off-resonant cavity photon, detuned from the atomic resonance by the frequency $\dca$, is absorbed and then re-emitted into the cavity field.  As expected from second-order perturbation theory, the cavity energy shift due to the single atom, located at position $\mathbf{r}_i$, is given as $\hbar |g^2(\mathbf{r}_i)| / \dca$ (this relation is exhibited in Eq.\ \ref{eq:e1pm} in the limit $|\dca| \gg |g(\mathbf{r})|$).  For many atoms in the cavity, this energy shift is additive, giving the Hamiltonian
\begin{equation}
H = \hbar \omega_c \hat{a}^\dagger \hat{a} + \hmech + \sum_i \hbar \frac{|g^2(\mathbf{r}_i)|}{\dca} \hat{a}^\dagger \hat{a},
\end{equation}
where $\omega_c$ is the cavity resonance frequency, $\hat{a}$ the cavity photon annihilation operator, the sum is taken over all atoms, and we omit some constant energy terms. The term $\hmech$ describes the dynamics of atomic motion.

We now consider two simple scenarios for the atomic ensemble.  First, we consider the atoms to be tightly confined in a harmonic trap, centered at position $z_0$, with vibrational frequency $\omega_z$ along the cavity axis, and neglect motion in the transverse directions. The position of atom $i$ is taken to be $z = z_0 + \delta z_i$ and thus $g(\mathbf{r}_i) = g_0 \sin(\phi_0 + 2 k \delta z_i)$ with $\phi_0 = k z_0$. Expanding to first order in the small Lamb-Dicke parameters $k \delta z_i\ll 1$, we then obtain \cite{purd10tunable}
\begin{equation}
H \simeq \hbar \left(\omega_c + N\frac{g_0^2}{\dca} \sin^2 \phi_0 \right) \hat{a}^\dagger \hat{a} + \hbar \omega_z \sum_i \hat{b}^\dagger_i \hat{b}_i + \hbar \frac{g_0^2}{\dca} \sin (2 \phi_0) \, \hat{a}^\dagger \hat{a} \, \sum_i k \delta z_i.
\label{eq:LDfirstorder}
\end{equation}
with $\hat{b}_i$ annihilating a phonon from the motion of atom $i$.  From the sum of the cavity coupling to each of the individual atoms, we identify a single collective atomic variable with which the cavity interacts.  Here, this collective variable is simply the center of mass $\Zcm = N^{-1} \sum \delta z_i$ of the ensemble.  For this simple mechanical setup, the center of mass motion is a normal mode of the system: a harmonic oscillator of frequency $\omega_z$, a mass $M = N m$ equal to that of the entire $N$-atom ensemble, and a harmonic oscillator length $\Zho = \sqrt{\hbar / 2 N m \omega_z}$.  Adjusting the cavity resonance frequency to $\omega_c^\prime = \omega_c + N (g_0^2 / \dca) \sin^2 \phi_0$, we thus obtain
\begin{equation}
\hamiom = \hbar \omega_c^\prime \hat{a}^\dagger \hat{a} + \hbar \omega_z \hat{b}^\dagger \hat{b} + \hbar \gom \left(\hat{b}^\dagger + \hat{b}\right) \hat{a}^\dagger \hat{a} + \hbath,
\label{eq:canonicalOMhami}
\end{equation}
Here, we write $\hmech = \hbar \omega_z \hat{b}^\dagger \hat{b} + \hbath$ where $\hat{b}$ annihilates a phonon from the center-of-mass collective mode, and $\hbath$ describes the remaining normal modes of the system.

We thus identify the canonical cavity optomechanical Hamiltonian for linear optomechanical coupling.  The single-photon/single-phonon optomechanical coupling strength is
\begin{equation}
\gom = N \frac{g_0^2}{\dca} \sin(2 \phi_0) \times k \sqrt{\frac{\hbar}{2 N m \omega_z}},
\label{eq:gomscaling}
\end{equation}
an expression which exhibits the scaling with atom number, atom-cavity detuning, and mechanical frequency, all of which can be varied broadly via straightforward modifications of the experimental system, even during a single experiment.

Building upon this example, we consider an ensemble split into several harmonic traps.  The cavity-chosen, linearly coupled, collective atomic variable is now a sum of the centers of mass of the sub-ensembles, weighted by their individual optomechanical coupling strengths  \cite{gupt07nonlinear,murc08backaction,schl11cooling}.  This setup is akin to having several solid-state membranes within a single resonator, coupled to one another via the cavity field.

\begin{figure}[t]
\begin{center}
\includegraphics[width=11.5cm]{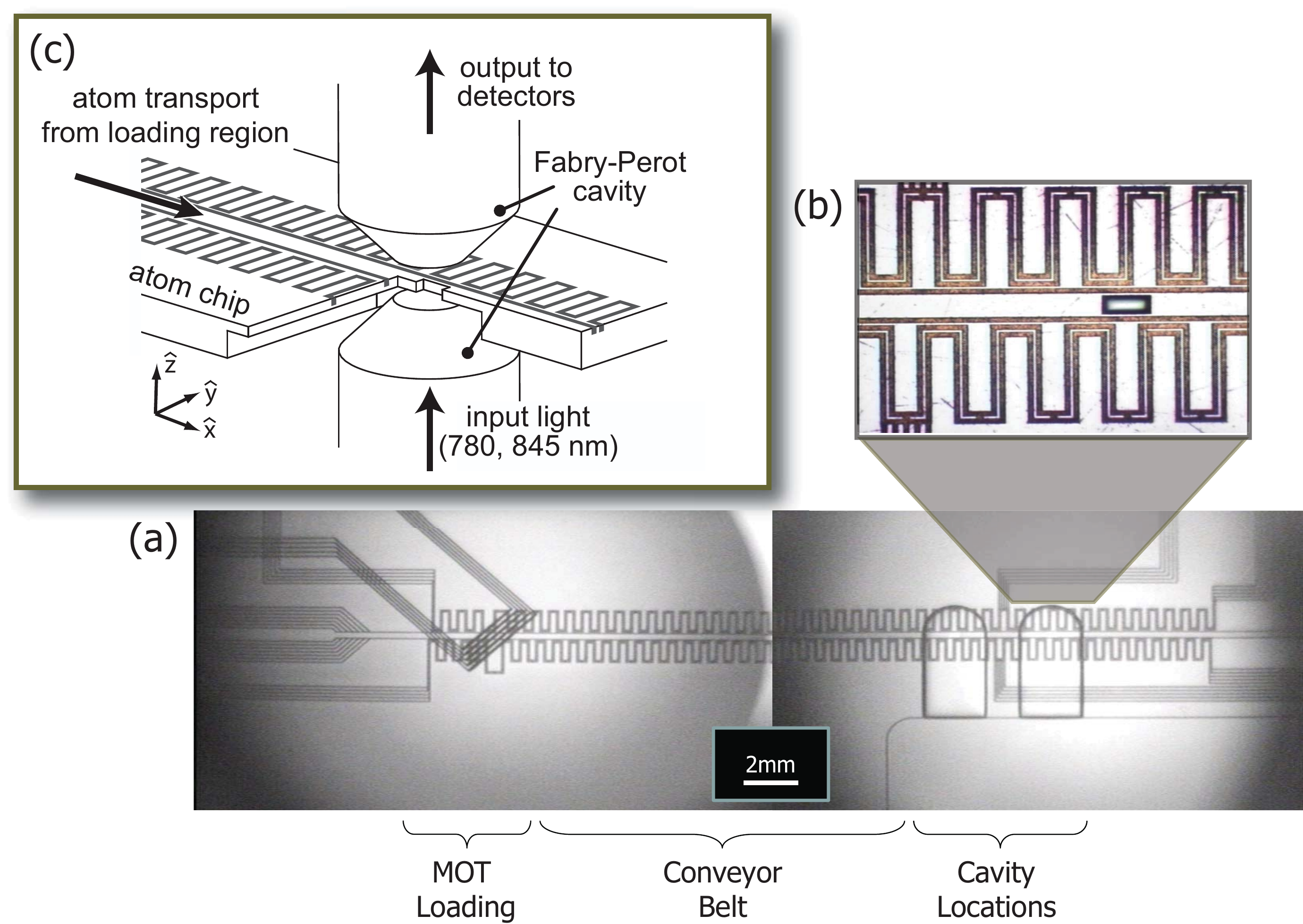}
\end{center}
\caption{The Berkeley atom chip experiment used to deliver an ultracold atomic ensemble into a high-finesse Fabry-\Perot\ cavity.  (a) A transmission image of the silicon-substrate atom chip shows inlaid copper wires on front and back surfaces as opaque lines.  These wires are used as electromagnets to generate the magnetic fields required to load atoms into a magneto-optical trap (left), transport them across the chip in a magnetic conveyor system (middle), and then confine them tightly and position them within the cavity region (right).  The tombstone-shaped shadows indicate regions where the atom chip is etched down to a thickness of just 100 $\mu$m.  (b) Within this thinned region, the atom chip is perforated to allow cavity light through the chip.  (c) Two curved mirrors are positioned above and below this perforation to form the Fabry-\Perot\ cavity, with a mirror spacing of about 250 $\mu$m.  Figure (c) is reproduced from Ref.\ \cite{purd10tunable}.}
\label{fig:atomchip}
\end{figure}

As with the membrane-based realizations of cavity optomechanics \cite{thom08membrane}, the linear optomechanical coupling strength varies with the equilibrium position of the mechanical element within the cavity.  This variation was demonstrated by the Berkeley group \cite{purd10tunable}.  In these experiments, a cold atomic gas of several thousand atoms was produced and translated into an optical resonator by means of a microfabricated atom chip (Fig.\ \ref{fig:atomchip}).  The chip provided for strong magnetic confinement of the gas, although not sufficiently strong to reach the Lamb-Dicke regime.  Stronger confinement was then provided by driving a TEM$_{00}$ mode of the optical cavity, at a frequency that was far red-detuned from the atomic resonance.  This strong cavity field created a one-dimensional optical lattice potential, within which the atomic gas occupied just two or three adjacent wells \cite{brah11mri}.  Each of these wells provides a near-harmonic trap, relevant to the derivation above.  Due to the small wavevector difference $q = k_p - k_t$ between the trapping light ($k_t = 2 \pi / (850 \, \mbox{nm})$) and probe light (near the $^{87}$Rb atomic resonance at $k_p = 2 \pi / (780 \, \mbox{nm})$), the optomechanical coupling strength within neighboring wells of the optical lattice was similar, with $\phi_0$ changing by 0.26 rad from well to well.  Broad variations in $\phi_0$ were achieved by varying the position of the magnetic trap.  The tuning of the optomechanical interactions was indicated both by the strong variation in the atoms-induced cavity resonance shift ($\omega_c^\prime - \omega_c$), and by the variation in the optomechanical frequency shift (Sec.\ \ref{sec:back action}).

\begin{figure}[t]
\begin{center}
\includegraphics[width=8cm]{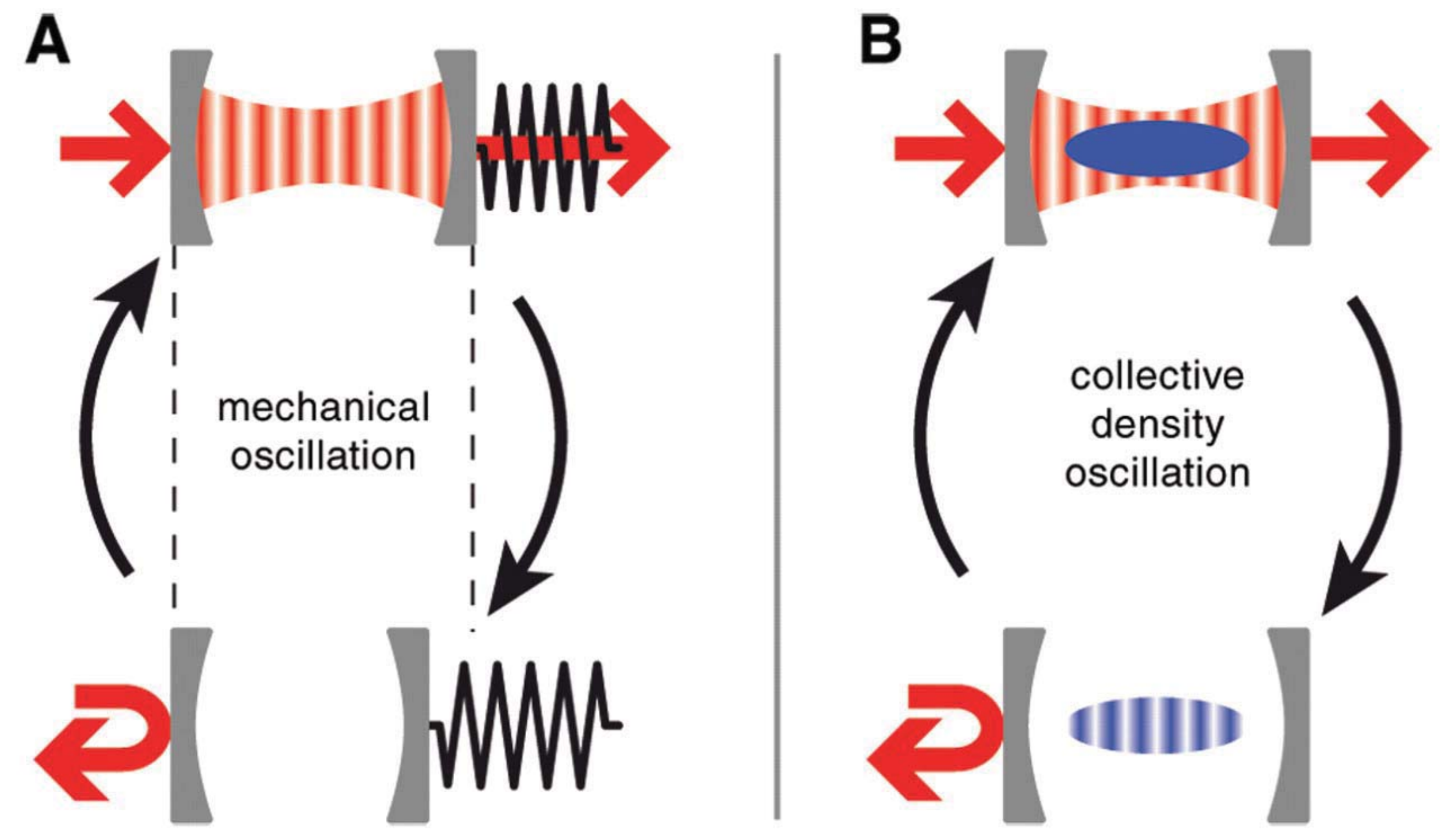}
\end{center}
\caption{The equivalence of cavity optomechanics realized with (a) moving mirrors or (b) cosine-mode density modulations within a Bose-Einstein condensate placed inside and coupled dispersively to the optical cavity.  Figure reproduced from Ref.\ \cite{bren08opto}.}
\label{fig:brennecke}
\end{figure}

A second simple initial state is a uniform, stationary, non-interacting Bose-Einstein condensed gas (Fig.\ \ref{fig:brennecke}) \cite{bren08opto}.  We emphasize that the assumptions of the gas being Bose condensed and non-interacting are not essential.  Bose-Einstein condensation ensures that density modulations at a fixed wavevector describe a normal mode of the system with a very sharp frequency response.  In comparison, density modulations of a collisionally thin non-degenerate gas dephase more rapidly.  Assuming the gas to be non-interacting allows us to neglect Bogoliubov transformations and the diminished structure factor for optically exciting phonons rather than free particles \cite{stam99phon}.  However, including such interaction effects presents no major difficulty.

With these assumptions, we recognize that, to lowest order, the spatial variation of $g^2(\mathbf{r}) \simeq g_0^2 \sin^2 k z = g_0^2 \left(1 - \cos 2 k z\right)/2$ couples the condensate to a cosine-function superposition of its $\pm 2 \hbar k$ momentum excitations, with the excitation energy $\hbar \omega_{2 k} = 2 \hbar^2 k^2/m$ equal to four times the recoil energy.  Using the Bose field operators $\hat{b}_0$ for the condensate and $\hat{b}_{1} = (\hat{b}_{+2 k} + \hat{b}_{- 2 k})/\sqrt{2}$ for the cosine-excitation mode where $\hat{b}_{\pm 2 k}$ are momentum-space field operators, and substituting $\hat{b}_0 = \sqrt{N}$, we approximate \cite{bren08opto}
\begin{equation}
H \simeq \hbar \left(\omega_c + N \frac{g_0^2}{2 \dca}\right) \hat{a}^\dagger \hat{a} + \hbar \omega_{2 k} \hat{b}_1^\dagger \hat{b}_1 + \sqrt{N} \hbar \frac{g_0^2 \sqrt{2}}{4 \dca} \left(\hat{b}_1^\dagger + \hat{b}_1\right) \hat{a}^\dagger \hat{a}
\end{equation}
where now the single-photon/single-phonon optomechanical coupling strength is $\gom = \sqrt{N} (g_0^2 \sqrt{2} / 4 \dca)$.

A similar analysis can be applied to the situation of a cold Fermi gas located within the cavity mode \cite{kana10fermi}.  In this case, the cavity field may be coupled to zero-sound modes of the degenerate Fermi gas.  This example illustrates the fact that neither Bose-Einstein condensation, nor quantum degeneracy, are necessary ingredients of most atoms-based cavity optomechanical systems.

\subsection{Sensing collective atomic motion}

Now we review research findings on cavity optomechanics garnered with this atomic-ensemble approach.  The experimental procedures utilized in this research differ from those of solid-state experiments.  For instance, whereas the solid-state experiments generally persist in a steady state, the mechanical element used in the atoms-based experiments -- the atomic gas -- is short-lived.  Thus, the gas must be prepared and positioned anew within the cavity before each measurement.  This leads to experimental variability, e.g.\ in the atom number, temperature, and position, that must be controlled, or at least accounted for, in order to compile repeated measurements.

In the current Berkeley experiments, this variability is assessed and adjusted for by active feedback.  For instance, variations in the atom number cause the atoms-shifted cavity resonance frequency $\omega_c^\prime$ to vary between measurements, and also to vary \emph{during} a measurement due to the ejection of atoms from their trap.  To accommodate this variability, the cavity-probe light is actively stabilized to a fixed detuning from the cavity resonance.

\begin{figure}[t]
\begin{center}
\includegraphics[width=\columnwidth]{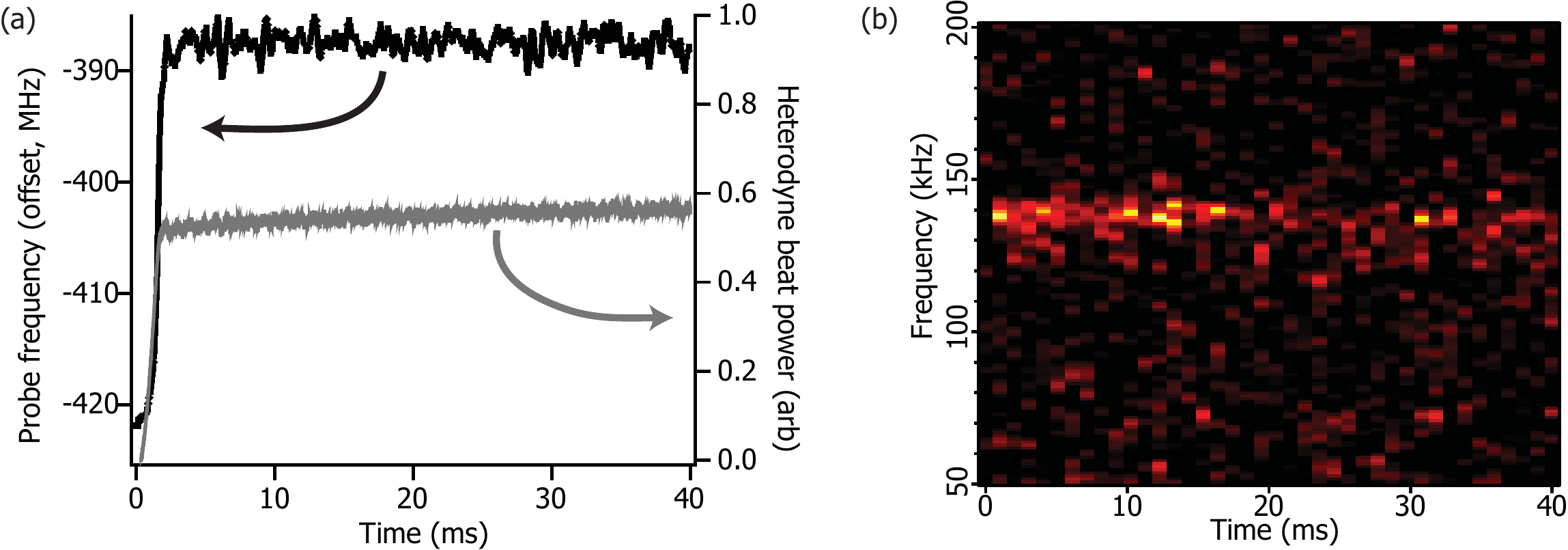}
\end{center}
\caption{Detecting motion in a newly prepared cold-atom optomechanics system.  (a) Shot-to-shot variations in the number of atoms trapped within the cavity makes the atom-shifted cavity resonance frequency $\omega_c^\prime$ uncertain at the start of an experiment.  To probe the cavity at a reliable detuning from the cavity resonance, the probe light is switched on first at a frequency far from the cavity resonance, and then active stabilization is used to tune it to the side of the cavity resonance.  The probe frequency (gray line, left axis) is seen to drift toward the resonance frequency and lock within about 1 ms.  The transmitted probe intensity, seen here as the power on a heterodyne detector (black line, right axis), stabilizes to a near-constant value.  The probe remains locked while atoms are slowly lost from the cavity, causing its frequency to drift slowly over a few MHz.  (b) Demodulating the heterodyne signal, and taking the Fourier transform of the demodulated signal within 1 ms time bins, reveals a signal peak at around 140 kHz, recording the motion of the mechanical element within the cavity.  The signal frequency and strength varies over 10's of ms as the atomic ensemble becomes increasingly perturbed by the constant probe.}
\label{fig:lockandprobe}
\end{figure}

The collective atomic motion is then sensed via the cavity field.  For example, a single record of the cavity transmission collected with the probe locked onto the side of the cavity resonance shows a sharp frequency modulation at the mechanical resonance of a harmonically bound atomic ensemble (Fig.\ \ref{fig:lockandprobe}).  This motional signal varies over the lifetime of the atomic ensemble, presumably due to light-induced displacements and heating of the ensemble, and finally disappears when the sample is fully depleted due to probe-induced heating.

\subsection{Back action onto the mechanical oscillator}
\label{sec:back action}

\begin{figure}[t]
\begin{center}
\includegraphics[width=11.5cm]{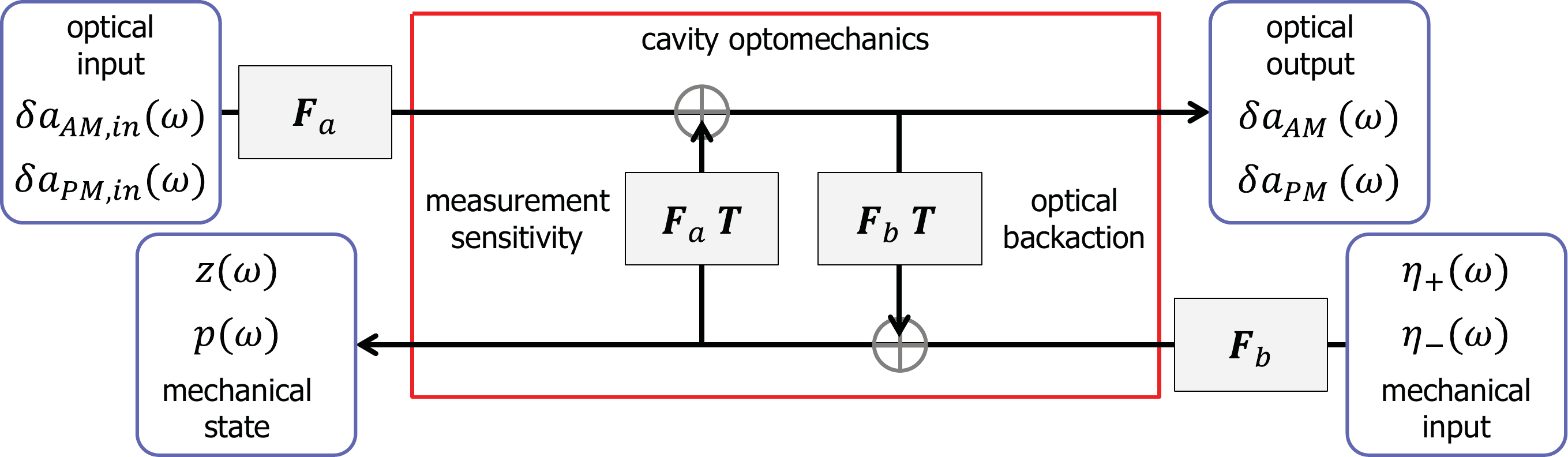}
\end{center}
\caption{The interactions between the optical field and the mechanical oscillator in cavity optomechanics are represented as a closed-loop feedback system, with the approximation that optical field is linearized about a steady coherent-state field with amplitude $\bar{a}$.  The system is driven by optical inputs, the amplitude (AM) and phase (PM) modulation at the frequency $\omega$, and mechanical inputs, $\eta_\pm(\omega)$ representing force modulations and fluctuations due to mechanical dissipation.  The feedback system relates these inputs to the optical and mechanical outputs according to the closed-loop, ponderomotive gain spectrum.  Elements in the feedback system include the cavity field and mechanical dynamical responses, summarized by transfer matrices $\mathbf{F}_a$ and $\mathbf{F}_b$, respectively, and the linear optomechanical coupling $\mathbf{T}$, which is proportional to $\gom \bar{a}$.  For example, the maximum closed-loop optical-to-optical response is proportional to the optomechanical cooperativity $4 \gom^2 \nbar / \kappa \gamma$, with $\nbar = |\bar{a}|^2$, and the linewidths quantify the strength of the optical cavity (via $\mathbf{F}_a$) and mechanical oscillator (via $\mathbf{F}_b$) resonant responses.  The figure is adapted from Ref.\ \cite{bott12theory}.}
\label{fig:closedloop}
\end{figure}

The sensitivity of the cavity field to collective atomic motion can be thought of as one link in a closed-loop response connecting the optical and mechanical inputs and outputs, schematically represented in Fig.\ \ref{fig:closedloop}.  A formal and quantitative representation of this closed-loop schematic is presented in Ref.\ \cite{bott12theory}.

Let us make the \emph{linearization approximation,} wherein we take the cavity field to be $\hat{a} = e^{- i \omega_p t} \left(\bar{a} + \delta \hat{a}\right)$, where $\omega_p$ is the probe (or pump) optical frequency, $\bar{a}$ is the coherent-state amplitude of the driven cavity field, and we neglect terms quadratic in the field fluctuations $\delta \hat{a}$.  This linearization approximation is valid in the non-granular regime where $\gom$ is small compared to the cavity linewidth \cite{gupt07nonlinear,murc08backaction}.  With this approximation, the canonical cavity optomechanical Hamiltonian (Eq.\ \ref{eq:canonicalOMhami}) becomes
\begin{equation}
-\hbar \Delta \hat{a}^\dagger \hat{a} + \hbar \omega_z \hat{b}^\dagger \hat{b} + \hbar  \gom |\bar{a}|^2 \left(\hat{b}^\dagger + \hat{b}\right) + \hbar \bar{a} \gom \times \left(\hat{b}^\dagger + \hat{b}\right)  \left( \delta\hat{a}^\dagger + \delta\hat{a}\right)
\label{eq:linearizedhom}
\end{equation}
where we have assumed without loss of generality that $\bar{a}$ is real.  The energy of the cavity is now measured with respect to the probe frequency, with $\Delta = \omega_p - \omega_c$.  The optomechanical interaction term proportional to $|\bar{a}|^2$ describes a constant force on the oscillator, and may be absorbed by shifting its equilibrium position. The linearized optomechanical coupling strength is now $\gom \bar{a}$, and the interaction contains only bilinear terms.

Within this linearized picture, we state that the motion of the oscillator $z(\omega)$ translates to a modulation of the cavity field (generally in both its amplitude and phase quadratures) via a transfer function $\mathbf{F}_a \mathbf{T}$ that is a product of two matrices: an optomechanical coupling matrix $\mathbf{T}$ that is proportional to $\gom \bar{a}$, and a cavity conditioning matrix $\mathbf{F}_a$ that describes the evolution of amplitude and phase modulations in the cavity.  The latter matrix depends on the detuning $\Delta$ of the cavity probe from the cavity resonance, and captures the fact that the motion is recorded on the phase quadrature if the cavity is probed on its resonance, and in a combination of amplitude and phase quadratures if probed off resonance.

Simultaneously, the cavity field influences the motion of the mechanical oscillator.  This influence is summarized by the transfer function $\mathbf{F}_b \mathbf{T}$ connecting the two quadratures of field fluctuations to the two quadratures of mechanical fluctuations.  This connection emphasizes the fact that, in quantum mechanics, the act of measurement is necessarily influential on the object being measured.

\begin{figure}[t]
\begin{center}
\includegraphics[width=6cm]{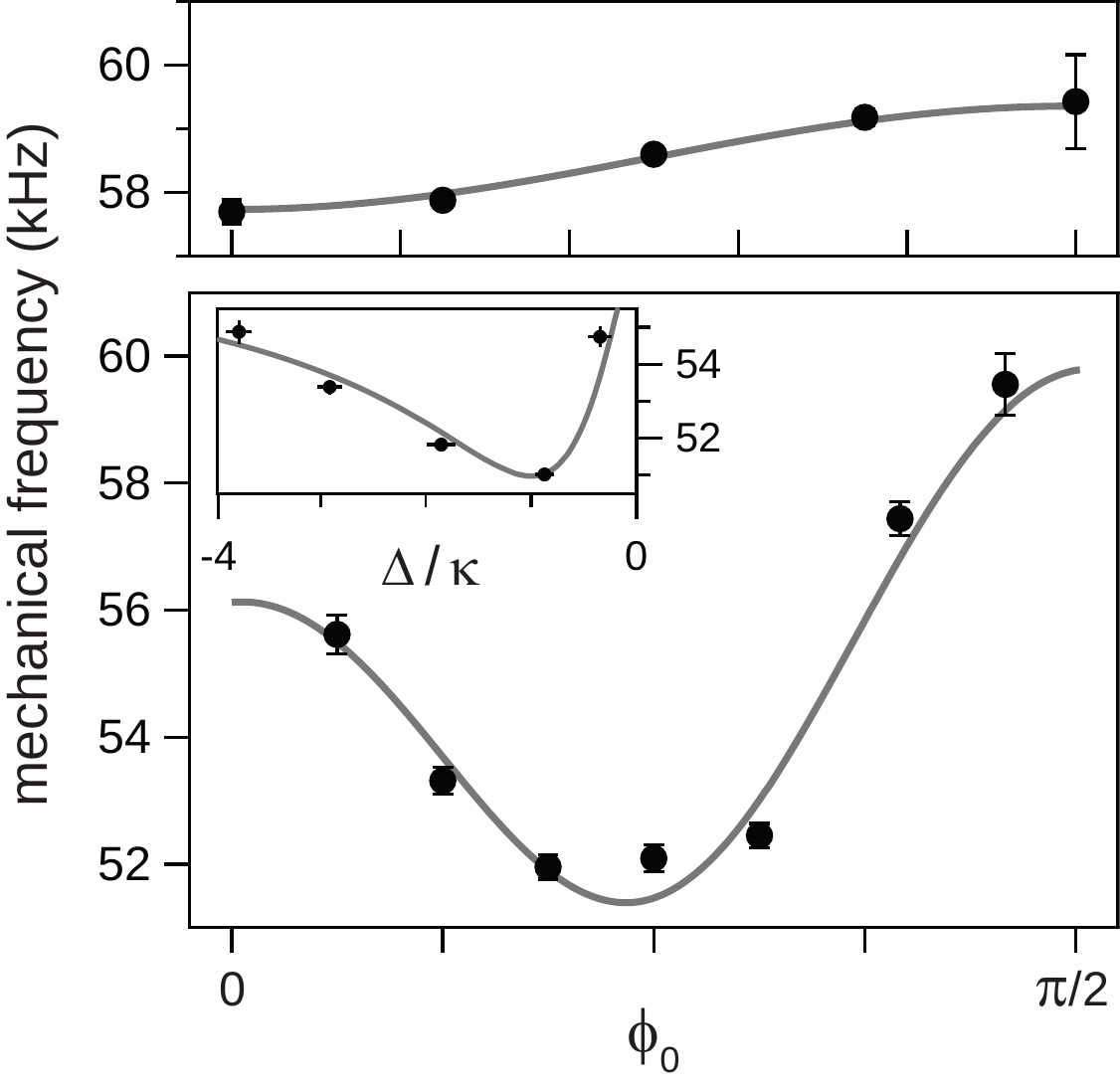}
\end{center}
\caption{Optomechanical frequency shifts measured with an atomic ensemble trapped at various positions along the cavity axis.  Top: The mechanical
oscillation frequency $\omega_z$ determined via parametric heating reveals the static frequency shift from quadratic optomechanical coupling; here,
$\Dca/2\pi\!=\!20.1 \, \mbox{GHz}$, $\omz/2\pi\!=\!58.5 \,
\mbox{kHz}$ and $\bar{n}\!=\! 0.8$. Bottom: the oscillation
frequency for the linearly coupled mode is derived from the cavity
transmission intensity spectrum.  The deviation between the frequencies in the top and bottom figures quantifies the optomechanical frequency shift due to linear optomechanical coupling.  This frequency shift is maximal in the case of maximal linear coupling ($\phi_0 = \pi/4$).  Here, $\Dca/2\pi \!=\! 40 \,
\mbox{GHz}$, $\omz/2\pi\! =\! 58.9 \, \mbox{kHz}$, $\nbar \!=\! 3.5$
and $N\! =\! 3750$. Inset shows this frequency at $\phi_0=\pi/4$ for
varying detuning $\Delta$ from the cavity resonance.  Solid lines
show predictions calculated with no free parameters.  Figure reproduced from Ref.\ \cite{purd10tunable}.}
\label{fig:frequencyshift}
\end{figure}

The effects of such back action may be described as either \emph{coherent} or \emph{incoherent}, based on whether or not quantum fluctuations play a role.  One form of coherent back action is the optical spring effect, in which the mechanical oscillation frequency is shifted by the adiabatic response of radiation pressure to the moving oscillator \cite{shea04spring,corb06spring,corb07gram}.  The optomechanical frequency shift was analyzed by the Berkeley group \cite{purd10tunable}.  Positioning the harmonically trapped atoms at the location of maximum linear optomechanical coupling, the frequency shift was measured as a function of probe-cavity detuning $\Delta$, and found to be in quantitative agreement with a no-free-parameter theoretical prediction (Fig.\ \ref{fig:frequencyshift}), demonstrating the appropriateness of using cavity optomechanics to describe mechanical effects of atomic ensembles in cavities.

\begin{figure}[t]
\begin{center}
\includegraphics[width=\columnwidth]{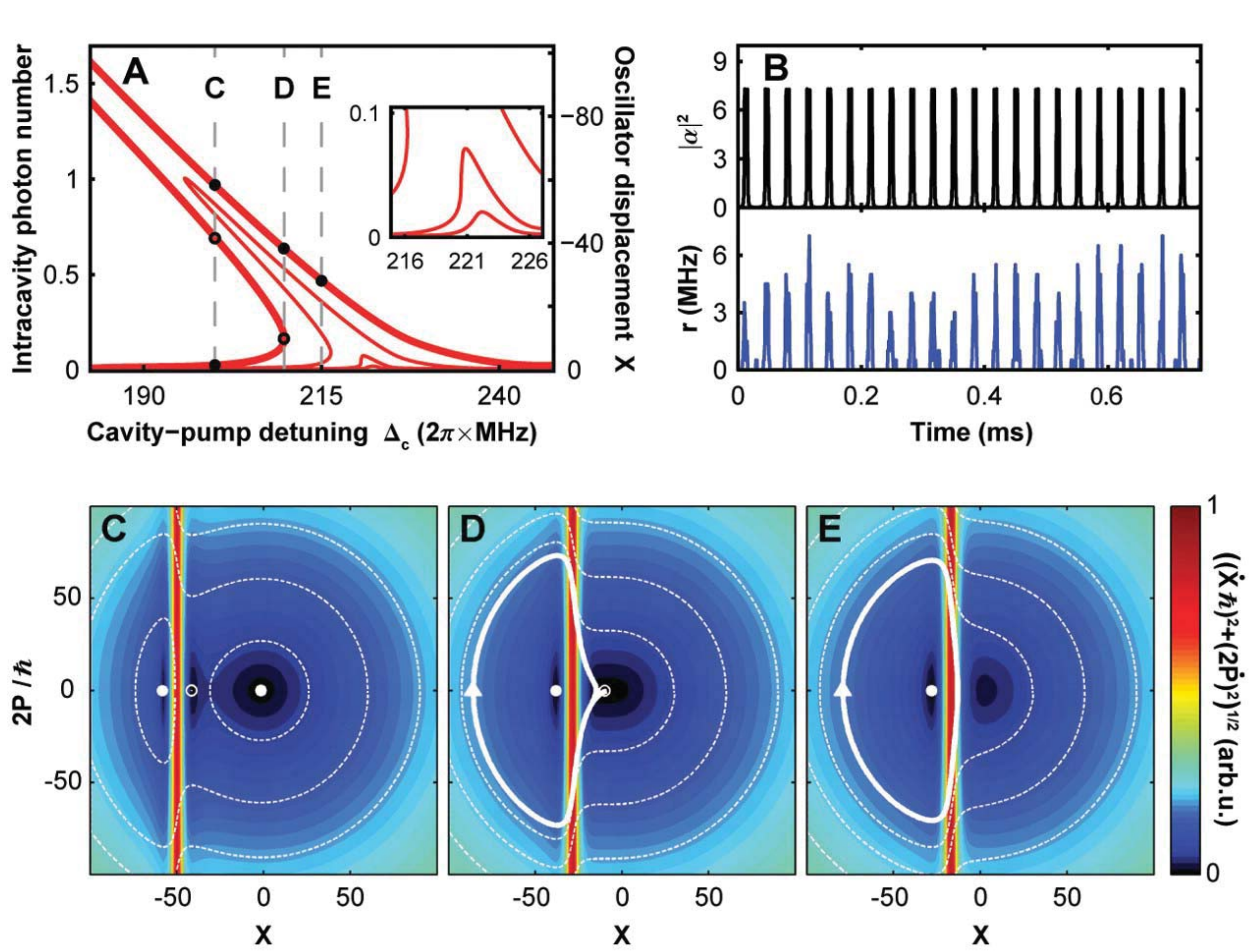}
\end{center}
\caption{Large scale motion of the mechanical oscillator leads to complex mechanical and optical dynamics.  (A) The cavity optomechanical system is operated at strong input probe power, reaching deep into the regime of optomechanical bistability.  (B) In this regime, the cavity output (top: theory, bottom: experiment) is a nearly periodic sequence of sharp emission bursts.  (C-E) The mechanical trajectory corresponding to these bursts is shown on a phase-space plot.   The mechanics evolve in a dark cavity until the cavity is tuned briefly into resonance with the probe, providing a strong radiation pressure impulse to the mechanical element and a burst of light through the cavity.  This pulsed force occurs twice per mechanical cycle.  Figure reproduced from Ref.\ \cite{bren08opto}.}
\label{fig:zurichlargescale}
\end{figure}

The optomechanical frequency shift describes the adiabatic dynamic back action of the cavity field upon the mechanical oscillator in the limit of small oscillation amplitude, where the radiation pressure force varies linearly with the position of the oscillator.  For larger mechanical displacements, when the cavity resonance frequency is made to vary by an amount comparable to the cavity linewidth, the oscillatory mechanical dynamics become more complex.  Such large-scale motion, represented by periodic spikes in the power transmitted through the driven optical cavity, was observed by the \Zurich\ group (Fig.\ \ref{fig:zurichlargescale}) \cite{bren08opto}.  Their data also display strong optomechanical frequency shifts \cite{ritt09apb}.

\begin{figure}[t]
\begin{center}
\includegraphics[width=8cm]{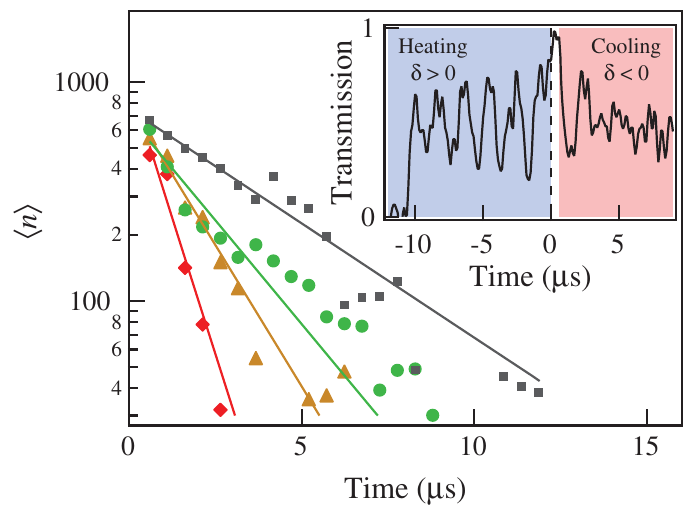}
\end{center}
\caption{Real-time observation of optomechanical cooling of an atomic ensemble.  Inset: The cavity is driven first with a blue-detuned probe, to excite large-amplitude motion of the intracavity ensemble, and then with a red-detuned probe, to observe the decay of this motion due to cavity cooling.  A single record of the cavity transmission is shown.  Main figure: The mean phonon occupation number is derived from the cavity transmission spectrum.  The cooling rate is shown to increase with probe intensity (black squares to red diamonds trend from low to high optical scattering rates).  Figure reproduced from Ref.\ \cite{schl11cooling}.}
\label{fig:MITcooling}
\end{figure}

A second coherent back action effect is cavity cooling.  Experimental studies of cavity cooling with atomic ensembles were carried out with gases at the 100-$\mu$K-range temperatures reached by laser cooling.  To observe directly the coherent damping of collective atomic motion, the researchers excited large scale motion, using cavity-induced amplification from a cavity probe that was blue-detuned from the cavity resonance, before switching the probe frequency to the red of the cavity resonance (Fig.\ \ref{fig:MITcooling}).  It was confirmed that the mechanical damping rate from cavity cooling, proportional to $\gom^2$, indeed scaled linearly with the atom number (see Eq.\ \ref{eq:gomscaling}), demonstrating that cavity cooling is a collective optical effect \cite{schl11cooling}.

It is important to point out that, in contrast with solid-state optomechanics for which cavity cooling is an essential means to bring solid-state objects toward their mechanical ground state (except for the highest-frequency oscillators that reach the ground state at dilution-refrigerator temperatures \cite{ocon10singlephonon}), atomic ensembles can be cooled to the quantum regime of motion by evaporative cooling to micro- or nanokelvin-range temperatures.  Indeed, for such ensembles, cavity-induced diffusive heating generally dominates cavity cooling, since the equilibrium temperatures reached by cavity cooling, on the order of $\hbar \kappa / k_B$, are typically higher than the ``base temperatures'' reached by evaporative cooling.  Nevertheless, cavity cooling may be advantageous for gases that cannot be cooled effectively by evaporative cooling, for example, gases of complex atoms or of molecules which suffer from large inelastic collision rates.

\subsection{Observations of radiation pressure shot noise}

Incoherent back action is represented by the quantum fluctuations of the radiation pressure force, representing the back action of a continuous quantum measurement of the position of the mechanical oscillator.  Considering the linear optomechanical coupling Hamiltonian $\hamiom$ (Eq.\ \ref{eq:canonicalOMhami}), the force on the mechanical oscillator can be taken as
\begin{equation}
\hat{f} = - \frac{\partial}{\partial \hat{Z}} \hamiom = - \frac{\hbar \gom}{\Zho} \hat{a}^\dagger \hat{a}
\end{equation}
Fluctuations in this force lead to diffusion, leading to an increase in the energy of the mechanical object.  For an undamped harmonic oscillator, the energy increases at a rate proportional to the spectrum of force fluctuations -- or equivalently, according to the above expression, the spectrum of photon number fluctuations -- at the mechanical frequency.

In the absence of technical fluctuations, a linear cavity driven by a coherent-state input, detuned by $\Delta$ from the cavity resonance, has a (two-sided) spectrum of photon number fluctuations given as
\begin{equation}
\snn(\omega) = 2 \nbar \frac{\kappa}{\kappa^2 + (\Delta + \omega)^2}
\label{eq:snn}
\end{equation}
where $\nbar$ is the average intracavity photon number.  Unlike in free-space, where the shot-noise spectrum is white, within a cavity the photon shot noise is accentuated at the frequency $\omega = - \Delta$ by the cavity resonance.  These intracavity fluctuations can be regarded as coming from the white field fluctuations of the cavity input.  At the cavity resonance, this input noise enters the cavity where it beats with the coherent-state field to produce strong photon number fluctuations. Away from the cavity resonance, this input noise is predominantly reflected and does not enter the cavity, causing photon number fluctuations to be relatively suppressed.

It is no accident that the photon shot-noise spectrum also quantifies the sensitivity of a cavity-based position measurement of the mechanical oscillator.  Variations $\delta Z$ in the position are seen via variations $\gom \delta Z / \Zho$ in the cavity resonance frequency.  To first order, these lead to variations in the cavity electric field by the amount
\begin{equation}
\delta E = E_0 \frac{i \gom}{\kappa - i \Delta} \frac{\delta Z}{\Zho}
\end{equation}
where $E_0$ is the cavity field strength within the cavity with the oscillator at rest (so that $\bar{n} \propto |E_0|^2$), and we assume $\omega_z \rightarrow 0$ for simplicity.  In this limit, the signal strength, taken as $|\delta E|^2$, is indeed proportional to the photon shot noise spectrum $\snn$.

The cold-atoms approach has provided the first direct observations of radiation pressure shot noise in cavity optomechanics.  These observations are enabled by the fact that the atoms are so well isolated from their environment, that  radiation pressure fluctuations cause large accelerations of light mechanical objects, and that these forces are accentuated by the strong polarizability of atoms near their resonance frequency.

Radiation pressure shot noise was observed in four separate measurements by the Berkeley group: (M1) bolometric detection measured by atom loss, (M2) the observation of ponderomotive squeezing, (M3) the observation of ponderomotively amplified shot-noise fluctuations, and (M4) measurement of the heat flux onto the mechanical element due to force fluctuations.

In measurement (M1), these shot-noise driven fluctuations were observed by the bolometric quantification of the diffusive heating rate of the atomic ensemble.  Here, one makes use of the finite depth of the trapping potential, which allows the heating rate to be quantified by the atom loss rate.  This loss rate was measured as a function of the probe-cavity detuning $\Delta$.  At constant intracavity probe power, the diffusive heating rate showed the line shape predicted in Eq.\ \ref{eq:snn}, and matched the magnitude predicted for shot-noise-driven heating within about 20\% \cite{murc08backaction}.

\subsection{Mechanically induced nonlinear optics: bistability, ponderomotive amplification, and ponderomotive squeezing}

Returning to our schematic representation of cavity optomechanics in Fig.\ \ref{fig:closedloop}, we understand that optical signals entering the cavity act upon the mechanical element, which then acts back onto the light.  As such, the mechanical oscillator mediates optical self-interactions, providing the cavity with a nonlinear optical response.

Consider the dc effect of this self-interaction.  A weak probe of the cavity reveals the cavity resonance frequency $\omega_c^\prime$.  At higher probe-light power, radiation pressure shifts the mechanical oscillator's equilibrium position, shifting the cavity resonance.  The cavity is therefore a non-linear optical element, showing a cavity resonance that varies with the light intensity.

For sufficiently strong probe light, the light-induced shift of the mechanical oscillator is sufficient to shift the cavity resonance by more than its linewidth.  Under these conditions, the optical cavity becomes dispersively bistable.  Bistability induced by optical radiation pressure was observed in early experiments on solid-state optomechanics, using a mg-scale mass, a low-finesse optical resonator, and an optical power of several Watts \cite{dors83bistability}.  In experiments using atomic ensembles, the higher cavity finesse and lower mechanical mass leads to cavity bistability at much lower powers, observed experimentally as low as 100 fW corresponding to an average intracavity photon number $\nbar = 0.05$ far below unity \cite{gupt07nonlinear,ritt09apb}.  In side-pumped cavity optomechanical systems, it is also possible to observe regimes where no stable atom-cavity configurations exist; rather, the dynamics may be described by limit cycles \cite{grie11limit}.  Cavity bistability in atomic cQED is also observed due to atomic saturation at high photon number ($\nbar \geq 100$) \cite{remp91bistab,saue04cqed}, corresponding effectively to the mechanical response of \emph{electrons} rather than of collections of \emph{nuclei} (where the atomic mass resides).  However, owing to the much longer coherence time of the center-of-mass rather than internal-state excitation of the atom, optomechanical bistability occurs at threshold photon numbers that are lower by orders of magnitude.

The nonlinear response of the cavity optomechanical system varies with frequency, according to the dynamical response of the cavity and mechanical systems.  For weak optical signals and in the nongranular regime (see Sec.\ \ref{sec:granular} for discussion of the granular regime), the linearized cavity optomechanical system represented in Fig.\ \ref{fig:closedloop} allows one to define a closed loop optical gain relating the cavity input signals to the intracavity field \cite{bott12theory}.  The cavity is thus regarded as a ponderomotive amplifier.  The amplification spectrum has been measured in solid-state optomechanical systems using input optical drives that greatly exceed the optical shot noise level, so that the response to the deliberately modulated radiation pressure dominates over that of other mechanical and optical perturbations \cite{mari10signature,verl10}.  The same closed-loop optical response is observed also in experiments on optomechanically induced transparency \cite{weis10omit,safa11omit}, albeit by focusing on different optical inputs and outputs \cite{bott12theory}.

\begin{figure}[t]
\begin{center}
\includegraphics[width=7cm]{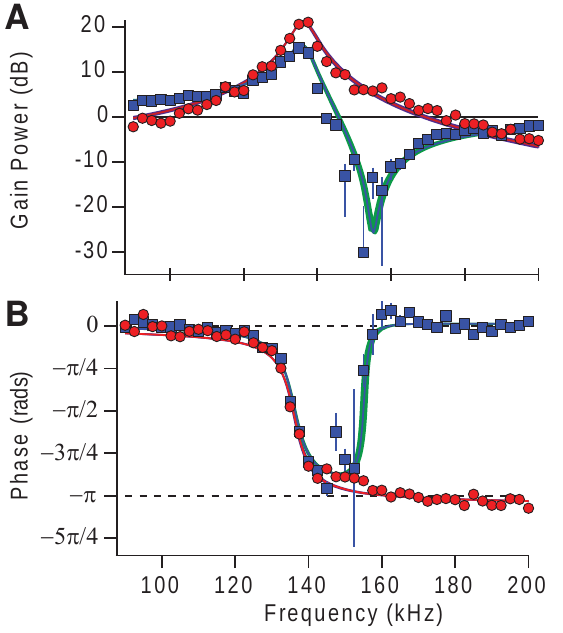}
\end{center}
\caption{Strong ponderomotive amplification is observed in a cavity optomechanical system pumped with just 36 pW of input power.  The gains for converting an AM input to either AM (blue squares) or PM (red circles) at the cavity output are complex numbers.  Here, (A) the power gains (square magnitude of the complex-valued gain) and (B) the phase shifts (arguments of the complex-valued gain) are plotted vs.\ the input modulation frequency.  The peak AM $\rightarrow$ PM gain occurs at the optically shifted mechanical resonance frequency, reaching a maximum power gain of 26 dB.  Above the mechanical resonance frequency, the out-of-phase mechanical response leads to a strong suppression of the AM output.  In the frequency region of strong suppression, one expects and indeed observes ponderomotive squeezing when the system is driven by a shot-noise-dominated probe field.  The agreement between measurement and theory (solid lines, with shaded region indicating systematic uncertainty) is remarkable. Figure reproduced from Ref.\ \cite{broo11pond}.}
\label{fig:gainspectrum}
\end{figure}

The ponderomotive amplification spectrum was measured using atomic ensembles (Fig.\ \ref{fig:gainspectrum}).  The cavity was pumped by a monochromatic coherent drive with the input power of around 36 pW, tuned below the cavity resonance, and then probed by placing single-tone amplitude modulation (AM) on the pump beam.  A heterodyne receiver at the cavity output was used to quantify the complex-valued gain for transducing the input AM tone to output AM and phase modulation (PM) at the cavity output.  At and below the optomechanically shifted mechanical frequency, the system showed strong amplification in both the AM and PM output quadratures, with a power gain as high as 20 dB.  This amplification results from the in-phase response of the mechanical oscillator to the radiation pressure modulation, which then feeds back so as to further amplify the cavity field modulations.  At frequencies above the mechanical frequency, where the oscillator responds out of phase with its force drive, the AM signal is suppressed by as much as 26 dB.  The remarkable quantitative match between measurements and the predictions of cavity optomechanics are, again, a testament to the utility of regarding collective atomic motion as equivalent to that of a solid-state mechanical oscillator \cite{broo11pond}.

Acting on the quantum fluctuations of the input field, the nonlinear suppression of input signals yields inhomogeneously squeezed light \cite{fabr94noise,manc94}.  Such ponderomotive squeezing differs from other methods of squeezing light in that it originates not from electronic motion, but rather from the collective motion of massive objects.  Ponderomotive squeezing has important consequences for quantum-limited and ``sub-quantum-limited'' detection of motion and forces, for example in gravity-wave observatories where it will modify the nature of squeezed light injected into the cavity optomechanical detector, and will require proper conditioning at the output of the detector in order to sense the squeezed, rather than the anti-squeezed, field quadratures \cite{kimb01ligo}.

Ponderomotive squeezing in an atoms-based cavity optomechanical system was observed following the aforementioned measurements of ponderomotive amplification simply by extinguishing the deliberate AM tone and allowing the system to be driven by shot-noise-dominated fluctuations.  In regions of ponderomotive suppression, the observed level of squeezing was small (about 1.5\% of shot noise) but statistically and systematically significant (errors at the level of 0.1\% of shot noise), and limited mostly by the poor quantum efficiency for photon detection in this system.

The detection of ponderomotive squeezing represents a direct observation of radiation pressure shot noise acting on the mechanical oscillator (measurement (M2)).  That is, light emitted from the cavity is a sum of the vacuum field fluctuations reflecting off the cavity mirror and of the intracavity optical field.  The reduction of optical quadrature fluctuations below their standard quantum limit confirms that the cavity field fluctuations are negatively correlated with those of the cavity input field by the back action of the noise-driven mechanical oscillator onto the cavity field.  In addition, in frequency regions of ponderomotive amplification, the cavity output spectrum was in quantitative agreement with that expected for a shot-noise driven optomechanical system (measurement (M3)).  Mechanical influences other than radiation pressure shot noise were constrained to have contributed no more than 3\% of the mechanical power spectrum.

\subsection{Sideband thermometry and calorimetry in quantum cavity optomechanics}

Ponderomotive squeezing represents a clear quantum signature of the ``opto'' portion of the cavity optomechanical system.  One signature for the quantum nature of the ``mechanical'' portion is the observation of asymmetric motional sidebands in light scattered off the mechanical oscillator.  This sideband asymmetry represents the fact that a quantum mechanical oscillator will more readily gain than lose energy.  That is, the sideband asymmetry reflects the fact that, exposed to classical (optical) force fluctuations (equal spectral densities for positive and negative frequencies), the $\nu$-phonon state will be excited by one phonon at a rate $R_+ \propto |\langle \nu+1 | Z | \nu\rangle|^2 \propto \nu + 1$ and de-excited at the smaller rate $R_- \propto |\langle \nu-1 | Z | \nu\rangle|^2 \propto \nu$.  Correspondingly, the powers in the emitted Stokes (red-shifted, $\propto \nu+1$) and anti-Stokes (blue-shifted, $\propto \nu$) sidebands are unequal.  The asymmetry is most pronounced for a ground-state harmonic oscillator: Since the oscillator cannot emit energy, the anti-Stokes sideband produced by its motion vanishes altogether.

The asymmetry between Stokes and anti-Stokes sidebands is readily observed for microscopic quantum systems, for example in the Raman spectra of molecular gases.  Sideband asymmetry is also observed in light scattered by single trapped ions \cite{died89}, single neutral atoms \cite{booz06cooling}, and also by chains of as many as 14 ions \cite{isla11qpt,monz1114} on the modes of collective motion cooled to near the ground state.  This telltale sideband asymmetry has been observed recently in two cavity optomechanical experiments: one involving a microfabricated solid-state mechanical oscillator \cite{safa12sideband}, and another involving the collective motion of an atomic ensemble \cite{brah12sideband}.

\begin{figure}[t]
\begin{center}
\includegraphics[width=11.5cm]{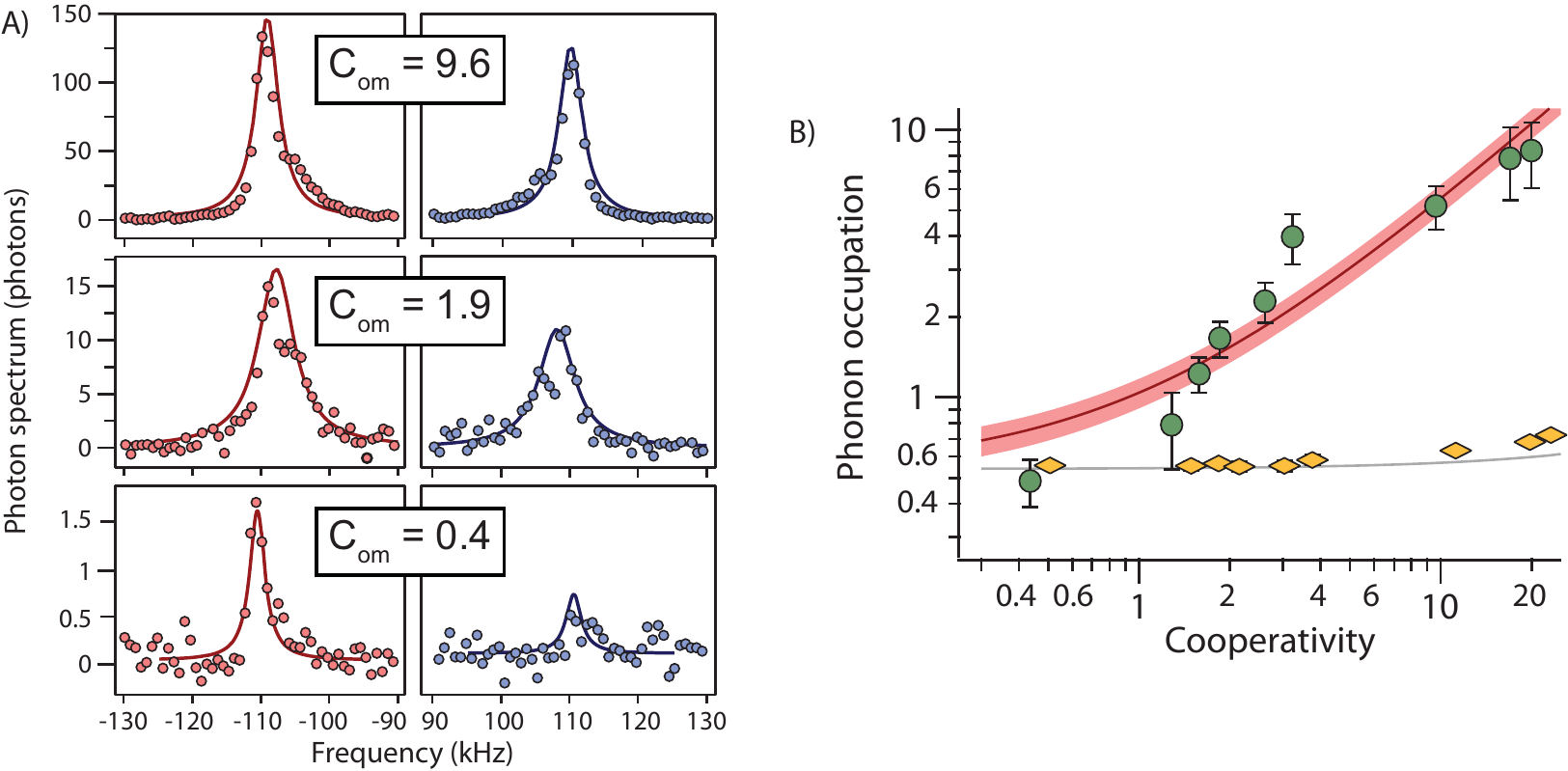}
\end{center}
\caption{Asymmetry between the Stokes (red) and anti-Stokes (blue) sidebands placed by the mechanical element on the cavity probe field reveals the quantization of collective atomic motion.  A: Sideband spectra are presented for probe power increasing from bottom to top.  B: The mean phonon occupation $\bar{\nu}$ determined by the sideband asymmetry (circles:data, shaded line:theory) is shown as a function of increasing probe power (quantified by the cooperativity $C_\mathrm{om} = 4 \gom^2 \bar{n} / \kappa \gamma$), and compared with the phonon occupation number expected at the measured temperature of the atomic ensemble (diamonds).  At low probe power, a 3:1 asymmetry is observed, corresponding to $\bar{\nu}=0.5$ and agreeing with the expected value at thermal equilibrium.  With increasing probe power, the sideband asymmetry diminishes, showing the collective atomic modes probed by the cavity to be selectively heated by measurement back action, while the remainder of the mechanical modes, acting as a large thermal bath, remain at nearly constant temperature.  Figure reproduced from Ref.\ \cite{brah12sideband}.}
\label{fig:sidebands}
\end{figure}

In the atomic-ensemble work, the cavity was probed with light at the cavity resonance and monitored in transmission by a heterodyne measurement.  Demodulating the heterodyne signal provided the sideband spectrum of the cavity field (Fig.\ \ref{fig:sidebands}).  For a weak cavity probe, the sidebands showed a 3:1 power ratio, indicating the mechanical system to have a steady-state average phonon occupation of $\bar{\nu} = 0.5$.  This was the value expected given the temperature at which the atomic ensemble was prepared by evaporative cooling, indicating that the collective mode (or a few modes) probed by the measuring the cavity-selected collective atomic variable was at equilibrium with the many other modes of motion within the ensemble.

For stronger cavity probes, the relative asymmetry of the sidebands diminished, indicating that $\bar{\nu}$ increased in response to the measurement.  In contrast, the overall temperature of the atomic gas, probed by time-of-flight methods, remained nearly constant, supporting the assumption that the measurement directly disturbs only the collective mode(s) being probed.

Further information is gleaned from the absolute, rather than the relative, asymmetry of the Stokes and anti-Stokes sidebands.  Every detected photon that is red-shifted by a frequency $\omega$ records the addition of an energy $\hbar \omega$ into the mechanical system, while blue-shifted photons record a similar subtraction of mechanical energy.  The cavity spectrum thus serves not only as a thermometer, but also as a heat-flux sensor, recording the spectrum of energy exchange between the optical and mechanical elements.  The observed spectra reported the net addition of energy into the mechanical system, at a flux that agreed quantitatively with the diffusive heating expected from quantum measurement back action (measurement (M4)).

\subsection{Cavity optomechanical effects of quadratic coupling}

As described in Sec.\ \ref{sec:collectivevariable}, for an atomic ensemble trapped in a single harmonic well, the linear optomechanical coupling strength (linear in $\Zcm$) varies according to the position of the gas within the standing-wave cavity mode.  Placing the ensemble at the node or antinode of the cavity field eliminates the linear coupling altogether, necessitating the inclusion of terms quadratic in the atomic displacement.  Continuing the expansion of Eq.\ \ref{eq:LDfirstorder} to second order in the Lamb-Dicke parameters, we obtain a quadratic optomechanical coupling term of the following form:
\begin{equation}
H_{\mathrm{quad}} = \hbar \gom \, k \Zho \,  \cos(2 \phi_0) \left(\frac{\Zcm^2}{\Zho^2} + \frac{\sigma^2}{\Zho}\right) \hat{a}^\dagger \hat{a}.
\end{equation}
The cavity frequency displays a quadratic response to the center-of-mass displacement, and also (dominantly) to the position variance $\sigma^2$ of the compressible atomic medium.  All normal modes of axial motion contribute to the position variance; hence, the quadratic optomechanical coupling senses, acts back upon, and potentially cools (by cavity cooling or by active feedback), all modes of motion in the atomic gas.

Quadratic optomechanical sensitivity is also achieved in membrane-based cavity optomechanics, although effects of this quadratic coupling have only been observed in the atomic-ensemble realization \cite{purd10tunable}.  One such effect is cavity nonlinearity and bistability induced by variation in the strain of the atomic gas: Cavity probe light either adds or detracts from the trap curvature, depending on the sign of the quadratic coupling term, causing the gas to either contract or expand, and thereby shifting the cavity resonance frequency.  The variation of the trap curvature with probe power also leads to a quadratic optomechanical frequency shift (Fig.\ \ref{fig:frequencyshift}).  Future work may explore phenomena such as ponderomotive amplification and squeezing induced by the quadratic coupling term.

\subsection{The granular regime of cavity optomechanics}

\label{sec:granular}

One of the most important contributions from the work on cavity optomechanics with atomic ensembles is the identification of a new, ``granular'' regime of cavity optomechanics, one in which the coupling between photons and phonons is significant at the single quantum level \cite{gupt07nonlinear}.  Quantitatively, the granular regime is reached when $\epsilon = \gom / \kappa > 1$.

To illustrate the implication of this relation, let us consider the strength by which an optical interrogation of the cavity measures the position of the mechanical element.  The transmission of a single photon through the cavity determines the cavity resonance frequency to within the cavity half-linewidth $\kappa$.  In turn, this knowledge determines the position of the mechanical element to within a position uncertainty of $\delta Z = \Zho (\kappa / \gom)$.  In the granular regime, $\delta Z < \Zho$, so that a single photon measurement is sufficient to collapse the position of the mechanical oscillator to an uncertainty smaller that the range of its zero-point motion.  The granular regime is thus an inherently strong quantum measurement regime.

Concomitantly, the granular regime is one in which the back action of a measurement is strong.  A single photon within the cavity exerts a radiation pressure force $f = - \hbar \gom / \Zho$ on the mechanical element.  Its residence time within the cavity is uncertain to within the cavity ring-down time, $\sim \kappa^{-1}$.  Thus, the impulse exerted by a single photon onto the mechanical element has an uncertainty of $\delta P = (\gom/\kappa) (\hbar / \Zho)$.  In the granular regime, $\delta P > \hbar / \Zho$, i.e.\ the measurement back action increases the uncertainty in the momentum of the mechanical element by an amount greater than its zero-point uncertainty.  Conversely, viewing the mechanical element as a detector for intracavity photons, the granular regime is one in which single photons passing through the cavity are strongly measured by imparting distinguishable changes in the mechanical state.

The reported experiments on cavity optomechanics with atomic ensembles have approached the granular regime, with the highest reported value of the granularity parameter being $\epsilon = 1.5$ in the work of Gupta \emph{et al.}\ \cite{gupt07nonlinear}.  In contrast, in solid-state cavity optomechanical systems, the granularity parameter is typically much smaller, below $10^{-2}$ in the recent work on ``optomechanical crystals,'' \cite{eich09crystal}, and several orders of magnitude lower still in other experiments.

It is unclear how the range of optomechanical effects obtained in the nongranular regime, such as those discussed previously in this Section, will be modified in the regime of stronger optomechanical coupling.  The common linearization approximation that is applied to describe those effects is clearly invalid in the granular regime.  Several theoretical works have begun to consider this regime, pointing to new features such as strong photon antibunching \cite{rabl11blockade}, cavity absorption and emission spectra dressed by several mechanical sidebands \cite{rabl11blockade,nunn11single,liao12}, unstable dynamics driven by strong radiation pressure fluctuations \cite{ludw08instability} and with distinct quantum signatures \cite{qian11instability}, photon bursts related to optomechanical bistability \cite{kron12fullstat}, and entanglement generated in an optical interferometer in which photons on one arm are effectively measured by the displacement of a mechanical element \cite{hong11interferometer}.  The last work relates to earlier proposals to measure gravity-induced decoherence of massive objects by entangling their motion with single photon states \cite{mars03mirror}, demonstrating that the granular regime is required for such investigations.  Many of these theoretical predictions rely on realizing the granular and resolved-sideband regime simultaneously, posing a clear, but attainable, challenge for future experiments.

\section{Cavity optomechanics of a side-pumped ensemble}
\label{sec:sidepumped}

Now we reconsider two experimental configurations discussed in Sec.\ \ref{sec:brillouin}:  an atomic ensemble within a Fabry-\Perot\ cavity pumped with a transversely oriented standing-wave field (Fig.\ \ref{fig:evenodd}), and within a ring-cavity pumped on one of its running-wave modes (Fig.\ \ref{fig:ringcavity}).  Unlike in the single-mode cavity-driven optomechanical systems discussed in the previous Section, here, the system is driven by a field external to an otherwise undriven cavity mode.  In the ring-cavity case, the undriven cavity mode (say, the right-going mode, as seen by the atoms) is one of the running wave modes, while the pump field is itself resonant within a second cavity mode (the left-going mode).  However, for the case of a strong pump field and only a weak occupation of the undriven cavity modes, we may neglect the cavity dynamics of the pumped mode and consider it simply as an externally imposed field. My aim in this Section is to treat these situations in the language of cavity optomechanics, both to supplement other descriptions of these systems in terms of, for example, realizations of quantum phase transitions, supersolidity, superradiance, and Brillouin instabilities, and also to set a precedent for similar experimental pursuits with solid-state cavity optomechanics.

\subsection{Cavity optomechanics of a side-pumped Fabry-\Perot\ resonator}

Consider the side-pumped Fabry-\Perot\ resonator studied by the MIT and \Zurich\ groups, shown schematically in Fig.\ \ref{fig:evenodd}.  We treat the standing-wave pump field initially as a finite-volume cavity mode with atom-cavity coupling strength $g_p(\mathbf{r}) = g_p \sin k x$, photon annihilation operator $\hat{a}_p$, and resonant frequency $\omega_p$.  The coupling strength to the Fabry-\Perot\ cavity is $g(\mathbf{r}) = g_0 \sin k z$.  In both cases we neglect the transverse variation of the field, assuming the atoms to be well confined at the intersection of the two fields.  There are three forms of coherent scattering: forward scattering into the pump field, forward scattering into the Fabry-\Perot\ field, and scattering of photons between the two fields.  Assuming a large optical detuning from the excited atomic state, the optomechanics of a \emph{single atom } in the cavity is given by the following Hamiltonian:
\begin{eqnarray}
H & = &  \hbar \omega_c \hat{a}^\dagger \hat{a} + \hbar \omega_p \hat{a}_p^\dagger \hat{a}_p + \hmech \\
& & + \hbar \frac{g_0^2}{\dca} \sin^2\left( k z\right) \,\, \hat{a}^\dagger \hat{a} + \hbar \frac{g_p^2}{\dca} \sin^2\left(k x\right)\,\, \hat{a}_p^\dagger \hat{a}_p + \hbar \frac{g_p g_0}{\dca} \sin\left(k x\right)\, \sin\left(k z\right) \left(\hat{a}^\dagger \hat{a}_p + \hat{a}_p^\dagger \hat{a}\right) \nonumber
\label{eq:sidepumped}
\end{eqnarray}
Now, we assume the pump field is a coherent state, allowing us to substitute $\hat{a}_p \rightarrow e^{-i \omega_p t} \bar{a}_p$.  Evaluating the cavity field in the probe's rotating frame, i.e.\ $\hat{a} \rightarrow e^{-i \omega_p t} \hat{a}$, we find
\begin{eqnarray}
H & = &  - \hbar \Delta \hat{a}^\dagger \hat{a} + \hmech \\
& & + \hbar \frac{g_0^2}{\dca} \sin^2\left( k z \right)\,\,\hat{a}^\dagger \hat{a} + \hbar \frac{G_p^2}{\dca} \sin^2\left( k x \right)  + \hbar \frac{G_p g_0}{\dca} \sin\left(k x\right) \sin\left(k z\right) \left(\hat{a}^\dagger + \hat{a} \right) \nonumber
\label{eq:sidepumpedlinear}
\end{eqnarray}
where $G_p = g_p \bar{a}_p$ and we dispense with the pump-field cavity altogether.

As in Sec.\ \ref{sec:collectivevariable}, to simplify the mechanical description of the atomic gas, let us specify the initial state of the gas and consider only small perturbations from that initial state.  Specifically, let us treat the case that the initial state is a non-interacting Bose-Einstein condensate.  Strictly speaking, with the cavity field empty, this condensate resides in the lowest Bloch state of the $x$-oriented optical lattice formed by the standing-wave probe field (orientations are shown in Fig.\ \ref{fig:evenodd}).  However, assuming this lattice to be weak, let us neglect it and consider the condensate to be spatially uniform; adapting our treatment to include this optical lattice is straightforward.  To match the spatial structure of the photon-exchange coupling term (last term in Eqs.\ \ref{eq:sidepumped} and \ref{eq:sidepumpedlinear}), we consider an excitation of the condensate with an annihilation operator of the form
\begin{equation}
\hat{b} = \frac{\hat{b}_{k(-\mathbf{x} + \mathbf{z})} + \hat{b}_{k(\mathbf{x} - \mathbf{z})} - \hat{b}_{k(\mathbf{x} + \mathbf{z})} - \hat{b}_{k(-\mathbf{x} - \mathbf{z})}}{2}
\end{equation}
where numerator contains momentum-space field operators, with $\mathbf{x}$ and $\mathbf{z}$ being unit vectors.  These momentum modes are all degenerate, with the excitation energy $\hbar \omega_M = \hbar k^2 / m$.

Let us also neglect the lattice potential $\hbar (g_0^2/\dca) \sin^2(k z) \hat{a}^\dagger \hat{a}$ formed by the cavity field itself, under the approximation that the pump field is far stronger than the cavity field.  For weak excitations atop the condensate, we thus obtain
\begin{equation}
H = -\hbar \Delta \hat{a}^\dagger \hat{a} + \hbar \omega_M  \hat{b}^\dagger \hat{b} + \hbar \frac{\sqrt{N} G_p g_0}{2 \dca} \left(\hat{b}^\dagger + \hat{b} \right) \left(\hat{a}^\dagger +  \hat{a} \right)
\end{equation}
This expression matches the canonical optomechanical Hamiltonian under the linearization approximation (Eq.\ \ref{eq:linearizedhom}), except in that, here, the mean cavity field is zero.  This connection between intracavity Brillouin instabilities and cavity optomechanics is discussed also in Ref.\ \cite{szir10noise}, focusing on aspects of ponderomotive amplification.

We are left with a very simple description of two harmonic oscillators coupled by a spring, a system whose normal modes are simply obtained from classical mechanics.  Let us focus on the optomechanical frequency shift in the limit of a weak cavity field.  If we approximate that the coupling frequency $\lambda = (\sqrt{N} G_p g_0)/(2 \dca)$ is small compared to the magnitude of the frequency difference $|\Delta| - |\omega_M|$, a condition satisfied in the experiments, we may apply second-order perturbation theory to determine the energy difference between the lowest phonon states, and obtain a frequency shift from $\omega_M$ by the amount
\begin{equation}
\delta \omega_M = \frac{\lambda^2}{\omega_M + \Delta}.
\end{equation}
As in the cavity-driven situation, the mechanical frequency is shifted upward for pump light that is blue-detuned from the cavity resonance, and shifted downward for red-detuned pump light.

Such optomechanical spring effects were observed recently by the \Zurich\ group, who describe the cavity-modified collective excitations of the Bose condensate as ``roton-like'' modes \cite{mott12roton}.  In the case of red detuning, as these modes are ``softened'' by the downward optomechanical frequency shift, indicating, as in liquid helium, the propensity of the fluid to solidify.  For red detuning and a sufficiently strong pump field ($|\lambda| > \sqrt{\omega_M (|\Delta| - \omega_M)}$), the mechanical spring constant changes sign, and the gas undergoes a dynamical instability.  As for an inverted harmonic oscillator, the ``position'' $\hat{b}^\dagger +  \hat{b}$ grows exponentially toward either large positive or large negative values; these correspond to the even or odd checkerboard (``solid'') patterns discussed in Sec.\ \ref{sec:brillouin} and shown in Fig.\ \ref{fig:brennecke}.  This instability thus breaks a $\mathbb{Z}_2$ symmetry of the initial mechanical state. The strength of the density modulation is stabilized by the energy terms ignored in our various approximations.

This optomechanical system also shows a strong analogy to the quantum Dicke model \cite{dick54}, which describes a collection of two-level atoms with equal electric-dipole coupling to a single cavity mode.  In this analogy, developed in Refs.\ \cite{nagy10dicke,baum10dicke}, the collective mechanical position operator $\hat{b}^\dagger + \hat{b}$ plays the role of the dipole moment operator, coupled to the electric field $(\hat{a}^\dagger + \hat{a})$.  The analogy breaks down for very large mechanical excitation, where rather than exhibiting saturation as one expects in the Dicke model, the mechanical system undergoes higher order Brillouin instability \cite{inou99super2,slam07super}.

The connection between cavity-mediated dynamics of a Bose-Einstein condensate and the Dicke model was anticipated by Refs.\ \cite{chen08dicke,morr08}, which treated the more general case of two bosonic modes coupled by light scattering involving a cavity-field photon and a driving-field (or a second cavity-field) photon. The model of Ref.\ \cite{chen08dicke} included also effects of self- and cross-interactions between the two bosonic modes, as well as a classical drive coupling the modes, as could be achieved in the cavity optomechanical system by illuminating the atoms both with a transverse pump field and also with a coherent cavity field.  The interesting variation between first and second order transitions in their work may, therefore, be observed in optomechanics experiments similar to those of the MIT and \Zurich\ groups.

One may ask whether the optomechanical instability described above and observed in experiments \cite{baum10dicke,blac03,baum11breaking} is a realization of the quantum phase transition exhibited by the zero-temperature quantum Dicke model, or of, perhaps, a classical phase transition exhibited in a finite-temperature system.  These situations are distinguished by whether quantum-mechanical or thermal fluctuations dominate near the phase transition.  Drawing on lessons from cavity optomechanics, one suspects that the mechanical system is beset by thermal fluctuations that originate from measurement-induced mechanical diffusion.  That is, the Fabry-\Perot\ field provides a constant measurement of the position of the mechanical oscillator.  This constant measurement implies the presence of force fluctuations; here, the relevant fluctuations originate from the radiation-pressure shot noise coming from the interference of vacuum fluctuations of the cavity field with the coherent-state field of the pump.  Indeed, recent measurements by the \Zurich\ group indicate that the system shows mechanical fluctuations much larger than the predicted zero-point fluctuations, with the disparity growing larger as the mechanical mode is softened when one approaches the phase transition \cite{note:zurich}.

Whether or not this system displays quantum phase transitions, it is highly significant as an example of an \emph{open} quantum system undergoing dynamical transitions under constant perturbation and measurement.  Predicted phenomena include complex long-term dynamics and dynamical multistability \cite{keel10dynamics,bhas12}, and an alteration of the critical exponents for the phase transition in an open system \cite{nagy11critical}.

\subsection{Cavity optomechanics in a pumped ring cavity}

For the ring-cavity setup (Fig.\ \ref{fig:ringcavity}), let the atom-cavity coupling strength to the running wave modes be $\gpm(z) \propto e^{\pm i k z}$, retaining only the spatial variation along the cavity axis.  We focus on the process whereby an atom backscatters light from one running wave mode to the other, giving a single-atom coupling strength of the form
\begin{equation}
\frac{\gminus^*(z) \gplus(z)}{\dca} \hat{a}_-^\dagger \hat{a}_+ + h.c.
\end{equation}
where $\hat{a}_{\pm}$ are the optical field operators for the respective cavity modes.  For the situation where one running-wave cavity mode (say the $+$ mode) is strongly driven and treated as a classical pump field with amplitude $e^{-i \omega_p t} \bar{a}_+$, this single-atom interaction term becomes
\begin{equation}
\frac{g_0^2 \bar{a}_+}{\dca} \left(e^{2 i k z} \hat{a}_- + e^{-2 i k z} \hat{a}_-^\dagger\right),
\label{eq:ringcavityoneatom}
\end{equation}
where $z$ is the atomic position, and $\bar{a}_+$ is real.

Let us again specify the initial state of the gas to be a uniform, stationary, and non-interacting Bose-Einstein condensate.  The spatially dependent optomechanical interaction term (Eq.\ \ref{eq:ringcavityoneatom}) couples this initial state to excitations at momenta $\pm 2 \hbar k$, with the well-defined energy $\hbar \omega_{2 k}$.  Summing over all $N$ atoms in the system and including the atomic excitation energy yields the following Hamiltonian:
\begin{eqnarray}
H &\simeq & - \hbar \Delta \hat{a}^\dagger \hat{a} + \hbar \omega_{2 k} \left(\hat{b}_{+2 k}^\dagger \hat{b}_{+ 2 k} + \hat{b}_{-2 k}^\dagger \hat{b}_{- 2 k}\right) \nonumber \\
& & + \hbar \sqrt{N} \frac{g_0^2 \bar{a}_+}{\dca} \left[ \left(\hat{b}_{+2 k}^\dagger + \hat{b}_{- 2 k}\right) \hat{a}_- + \left(\hat{b}_{+ 2 k} + \hat{b}_{-2 k}^\dagger\right) \hat{a}_-^\dagger\right]
\end{eqnarray}

In the case of the side-pumped Fabry-\Perot\ cavity, the checkerboard patterns favored by collective scattering emerge at fixed positions determined by the overlap of the two standing-wave fields.  In the case of the ring cavity, the position of the emergent standing wave pattern is not fixed.  To exhibit this fact, let us rewrite the optomechanical coupling term above in terms of the cosine and sine spatial modulation patterns; that is, we define $\bcos = (\hat{b}_{2 k} + \hat{b}_{-2 k})/\sqrt{2}$ and $\bsin =  (\hat{b}_{2 k} - \hat{b}_{-2 k})/(i \sqrt{2})$ and obtain the optomechanical coupling term as
\begin{equation}
\hbar \sqrt{2 N} \frac{g_0^2 \bar{a}_+}{\dca} \left[ \left(\bcos^\dagger + \bcos\right) \left(\hat{a} + \hat{a}^\dagger\right) + \left(\bsin^\dagger + \bsin\right) \left(\frac{\hat{a} - \hat{a}^\dagger}{i}\right)\right]
\end{equation}
The two independent quadratures of the cavity field are coupled to two independent harmonic oscillators.  The optomechanical frequency shift and the condition for optomechanical instability are treated similarly as for the Fabry-\Perot\ cavity setup discussed above.  The instability besetting the two mechanical modes is similar to that of a particle in a two-dimensional degenerate inverted harmonic oscillator potential.  The mechanical system spontaneously breaks a $U(1)$ symmetry by forming a density modulation at an offset position defined to within the modulation wavelength.  This symmetry breaking also selects the $U(1)$ phase of the backscattered running-wave cavity mode.

In the symmetry-broken phase, one cannot ignore the dynamical back action on the pumped cavity mode.  The phenomenology of nonlinear cavity optics in this system is quite rich, including cavity bistability, as observed by Els\"{a}sser \emph{et al.}\ \cite{elsa04}, and predicted regions of multi-stability and disconnected regions (``isolas'') of stable cavity operation that are flanked by unstable modes of operation \cite{chen10classical,stei11ring}.

\subsection{Optomechanics in side-pumped multi-mode cavities}

These experiments have given rise to theoretical proposals for continuous optomechanical systems with greater dynamical complexity.  Notably, Gopalakrishnan, Lev and Goldbart \cite{gopa09emergent} considered the low-energy states of an atomic gas that is placed within an optical cavity with a high degeneracy of cavity modes, such as a concentric or confocal Fabry-\Perot\ resonator, and illuminated by a standing wave of light.  The gas is susceptible to Brillouin instabilities into several cavity modes, each of which is supported by a distinct spatial pattern of the gas medium.  In this case, the gas may spontaneously adopt one of several distinct spatial (potentially ``supersolid'') distributions, as would be evident in the spatial mode of light emitted by the cavity.  In this sense, the ``crystallinity'' of the gas, rather than being externally imposed upon the atoms by an optical lattice of fixed geometry, now emerges due to the internal dynamics of the atoms-cavity system.  Alternately, different portions of the gas may support collective scattering into different cavity modes, producing a multi-mode optical output and a spatially modulated density pattern interrupted by dislocations or domain walls.  Applying this idea to the internal (spin) rather than the external (center-of-mass motion) degrees of freedom of the gas, this realization of the multi-mode Dicke model may produce spin-glass phases in a highly tunable, and essentially open, many-body quantum system \cite{gopa11prl,stra11dicke}.

In the language of cavity optomechanics, these proposals relate to a system in which several mechanical modes of the same medium are coupled to several optical modes.  Studies of multi-mode solid-state cavity optomechanics have only recently begun.  One expects both the atomic- and solid-state-based investigations of multi-mode optomechanical systems to yield new insight on optomechanical effects such as phonon lasing, amplification, synchronization and gain saturation; coherent energy transfer between mechanical modes; cavity-induced cooling; ponderomotive amplification and squeezing; etc.

\section{Future directions}

The clearest imperative for future research on cold-atomic cavity optomechanics is the exploration of optomechanics in the granular regime.  As discussed in Sec.\ \ref{sec:granular}, current experimental setups are sufficient to reach this regime.  Moderate increases in the mechanical trapping frequency and the cavity decay time will allow one to reach also the resolved sideband regime, so that distinctly nonlinear single-photon effects, such as photon antibunching \cite{rabl11blockade}, and other departures from weak-coupling cavity optomechanics may be observed.

Beyond this, theoretical studies indicate that novel optomechanical phenomena may arise by enriching the optomechanical system by adding multiple mechanical or optical modes.  Additionally, one may consider situations where the physical state of the mechanical medium, now considered as an interacting many-body system, affects or is affected by the cavity optomechanical interaction.  For instance, the equation of state of the gaseous mechanical medium changes when the gas undergoes a metal/superfluid to insulator transition.  This phase change may influence the cavity field, e.g.\, due to a change in the compressibility of the medium.  In turn, the cavity field, by changing the optical potential in which the gas resides, may itself drive the phase transition of the gas.  The self-consistent phase diagram of this atoms-cavity system has been considered for the case of the single-band Bose-Hubbard model, showing features that are distinct from those observed for atoms in free-space lattices \cite{chen09bistable,fern10}.  Another interesting intracavity medium to consider is the one-dimensional strongly interacting Bose gas, which shows a strong susceptibility to being ``pinned'' by weak periodic potentials, thereby enhancing the optomechanical response \cite{sun11oned}.  It will be interesting to examine how such many-body phases and phase transitions influence the dynamics of optomechanical systems.

Finally, several works have explored the analogy between the motional dynamics and the internal-state dynamics of atoms in a cavity.  By this analogy, new phenomena have been predicted to arise from the parametric coupling between spin ensembles and a single-mode cavity field, such as cavity magneto-optical bistability and spontaneous cavity birefringence, coherent amplification and damping of the collective spin, and the generation of inhomogeneously squeezed light \cite{brah10csod}.  It will be interesting to explore such phenomena experimentally with both atomic and solid-state spin ensembles.

\begin{acknowledgement}
The author is deeply grateful to his co-researchers on cQED with cold atoms, whose persistence, curiosity and keen insight led to the development of the optomechanics picture for describing the interactions of trapped gases with single-mode optical cavities.  This team includes Thierry Botter, Nathaniel Brahms, Daniel Brooks, Subhadeep Gupta,  Zhao-Yuan Ma, Kevin Moore, Kater Murch, Sydney Schreppler, and Thomas Purdy.  Additional contributions to the development of the experimental apparatus were made by Kevin Brown, Keshav Dani, Marilena LoVerde, and Guilherme Miranda.  I am thankful to T.\ Esslinger, H.J.\ Kimble, G.\ Rempe, H.\ Ritsch, V.\ Vuleti\'{c}, and C.\ Zimmermann for permission to use figures from their work, and also to A.\ Nunnenkamp and to P.\ Rabl for critical readings of the manuscript.  Financial support for our research was provided by the DARPA QuIST program, the NSF, the David and Lucile Packard Foundation, a critical seedling grant from DARPA through the AFOSR, and the AFOSR directly.
\end{acknowledgement}

\bibliographystyle{spphys}
\bibliography{allrefs_x2,optoreviewnotes}

\end{document}